\documentclass[manuscript]{aastex}
\usepackage{emulateapj5}
\usepackage{epsf}
\usepackage{onecolfloat}

\shortauthors{Stanek \etal}
\shorttitle{X-Ray Luminosity-Mass Relationship of Galaxy Clusters}

\newcommand{\lta}{\lesssim}
\newcommand{\gta}{\gtrsim}

\newcommand{\kmsmpc}{\,{\rm km}\,{\rm s}^{-1}\,{\rm Mpc}^{-1}}

\newcommand{\bdm}{\begin{displaymath}} 
\newcommand{\edm}{\end{displaymath}}
\newcommand{\beq}{\begin{equation}} 
\newcommand{\eeq}{\end{equation}} 
\newcommand{\bit}{\begin{itemize}} 
\newcommand{\eit}{\end{itemize}} 
\newcommand{\ben}{\begin{enumerate}} 
\newcommand{\een}{\end{enumerate}}
\newcommand{\bfi}{\begin{figure}[htb]} 
\newcommand{\bpfi}{\begin{figure}[p]}

\newcommand{\lnlf}{\hbox{${\ln}L_{15,0}$}}
\newcommand{\elf}{\hbox{L$_{15,0}$}}
\newcommand{\sigate}{\hbox{$\sigma_8$}}

\newcommand{\dm}{\hbox{$\sigma_{{\ln}M}$}}
\newcommand{\sigm}{\hbox{$\sigma_{{\ln}M}$}}
\newcommand{\sigM}{\hbox{$\sigma_{{\ln}M}$}}
\newcommand{\varM}{\hbox{$\sigma^2_{{\ln}M}$}}
\newcommand{\sigl}{\hbox{$\sigma_{{\ln}L}$}}
\newcommand{\sigL}{\hbox{$\sigma_{{\ln}L}$}}

\newcommand{\sigMT}{\hbox{$\sigma_{{\ln}M|T}$}}
\newcommand{\sigTL}{\hbox{$\sigma_{{\ln}T|L}$}}
\newcommand{\varTL}{\hbox{$\sigma^2_{{\ln}T|L}$}}
\newcommand{\varMT}{\hbox{$\sigma^2_{{\ln}M|T}$}}
\newcommand{\varML}{\hbox{$\sigma^2_{{\ln}M|L}$}}
\newcommand{\delML}{\hbox{$\delta^M_L$}}
\newcommand{\delMT}{\hbox{$\delta^M_T$}}
\newcommand{\delMLhat}{\hbox{$\hat\delta^M_L$}}
\newcommand{\delMThat}{\hbox{$\hat\delta^M_T$}}

\newcommand{\lnM}{\hbox{${{\ln}M}$}}
\newcommand{\lnL}{\hbox{${{\ln}L}$}}
\newcommand{\lnT}{\hbox{${{\ln}T}$}}

\newcommand{\lnLbar}{\hbox{$\overline{\lnL}$}}
\newcommand{\ficm}{\hbox{$f_{\rm ICM}$}}
\newcommand{\micm}{\hbox{$M_{\rm ICM}$}}
\newcommand{\mdelta}{\hbox{$M_\Delta$}}
\newcommand{\rdelta}{\hbox{$R_\Delta$}}
\newcommand{\xray}{\hbox{X-ray}}
\newcommand{\rhoc}{\hbox{$\rho_c$}}

\def \cf        {\hbox{\it cf.,} }

\def \etal      {\hbox{et al.} }
\def \dln       {d{\ln}}
\def \msol      {{\rm\ M}_\odot}
\def \h         {\hbox{$\, h$} }
\def \hinv      {\hbox{$\, h^{-1}$} }

\def \ergs      {\hbox{$\, {\rm erg~s}^{-1}$} }
\def \keV       {\hbox{$\,{\rm keV}$}}

\def \cgsflux   {{\rm\ erg\ s^{-1}\ cm^{-2}}}

\def \se        {\!=\!}
\def \sims      {\sim \!}
\def \ssim      {\! \sim \!}
\def \ssimeq    {\! \simeq \!}
\def \sequiv    {\! \equiv \!}
\def \spropto   {\! \propto \!}
\def\myputfigure#1#2#3#4#5
{\hskip0.03\textwidth\vskip#5pt
\makebox[0pt]{\hskip#2in
\includegraphics[width=#3\textwidth]{#1}}\vskip#4pt\hfill}     
\begin{document}

\title{The X-Ray Luminosity$-$Mass Relation for Local Clusters of Galaxies}

\author{R. Stanek\altaffilmark{1,5}, A.E. Evrard\altaffilmark{1,2,3}, H. B\"{o}hringer\altaffilmark{4}, P. Schuecker\altaffilmark{4}, B. Nord\altaffilmark{2} }

\altaffiltext{1}{Department of Astronomy and Michigan Center for Theoretical Physics, University of Michigan, 500 Church St., Ann Arbor, MI  48109}
\altaffiltext{2}{Department of Physics and Michigan Center for Theoretical Physics, University of Michigan, 450 Church St., Ann Arbor, MI  48109-1040}
\altaffiltext{3}{Visiting Miller Professor, Physics Department, University of California, Berkeley, CA 94720 } 
\altaffiltext{4}{Max-Planck-Institut f\"{u}r extraterrestrische Physik, D-85740 Garching, Germany}
\altaffiltext{5}{email: rstanek@umich.edu }

\begin{abstract} 

We investigate the relationship between soft \xray\ luminosity and mass for
low redshift clusters of galaxies by comparing observed number counts
and scaling laws to halo-based expectations of $\Lambda$CDM cosmologies.  We
model the conditional likelihood of halo luminosity as a log-normal
distributon of fixed width, centered on a scaling relation, $L \spropto
M^p\rhoc^s(z)$, and consider two values for $s$, appropriate for
self-similar evolution or no evolution.  Convolving with the 
halo mass function, we compute expected counts in redshift and flux
which, after appropriate survey effects are included, we compare
to REFLEX survey data.  Counts alone provide only an upper limit on the
scatter in mass at fixed luminosity, $\sigm < 0.4$. We argue that the
observed, intrinsic variance in the 
temperature--luminosity relation is directly indicative of
mass--luminosity variance, and derive $\sigm \se 0.43 \pm 0.06$ from
HIFLUGCS data.  When added to the likelihood
analysis, we derive values $p \se 1.59 \pm 0.05$, $\lnlf
\se 1.34 \pm 0.09$, and $\sigm \se 0.37 \pm 0.05$ for self-similar
redshift evolution in a concordance ($\Omega_m \se 0.3$,
$\Omega_\Lambda \se 0.7$, $\sigma_8 \se0.9$) universe.
The present-epoch intercept is sensitive to power spectrum
normalization, $L_{15,0} \spropto \sigate^{-4}$, and the slope is weakly
sensitive to the matter density, $p \spropto \Omega_m^{1/2}$. 
We find a substantially (factor $2$) dimmer intercept and slightly
steeper slope than the values published using hydrostatic
mass estimates of the HIFLUGCS sample, and show that a Malmquist bias
of the \xray\ flux-limited sample accounts for this effect.  In light
of new WMAP constraints, we discuss the interplay between parameters
and sources of systematic error, and offer a compromise model
with $\Omega_m \se 0.24$, $\sigma_8 \se 0.85$, and somewhat lower
scatter $\sigm \se 0.25$, in
which hydrostatic mass estimates remain accurate to $\ssim 15\%$.  
We stress the need for independent calibration of the L-M
relation via weak gravitational lensing. 

\end{abstract}



\section{Introduction}\label{sec:intro}

The counts, spatial clustering and bulk properties of the 
most massive halos in the universe offer a means to test cosmological physics
\citep{wang:98,  haiman:01, carlstrom:02, majumdar:03, battye:04,
  wang:04}.   Although only a handful of high redshift ($z \gta 1$)
clusters are currently known, the number will grow considerably in the
next five years, as a result of improved search techniques using
multi-band optical \citep{gladdersYee:05} combined with infrared
\citep{stanford:05} or \xray\ \citep{rosati:98, romer:00, mullis:05}
imaging.  Sunyaev-Zel'dovich (SZ) surveys, based on interferometric
\citep{Holder:00, loh:05}, or bolometric \citep{schwan:03,
  kosowsky:03, ruhl:04} approaches, will ultimately extend the reach
of the cluster population to $z \sim 3$.     

Since observational surveys do not select
directly on halo mass, interpreting the data requires a model that
relates observable bulk  features, such as temperature,
Sunyaev-Zel'dovich decrement, or \xray\ luminosity, to mass.  Given
sufficiently rich cluster samples, one can solve for parameters
describing the mass-observable relation along with cosmological
parameters in a simultaneous fashion \citep{levine:02, majumdar:03, limahu:04}.
Power-law relationships between bulk properties are expected on
dimensional grounds \citep{kaiser:86}, but 
the scatter about mean mass-observable relationships, a crucial element of
self-calibration exercises \citep{levine:02, limahu:05}, is presently poorly
understood.   

In this paper, we use counts and scaling relations of low
redshift clusters in the HIFLUGCS survey \citep{reiprich:02}  to investigate
the statistical relationship between \xray\ luminosity and total halo
mass.   The method relies on numerical simulations for calibration of
the space density of massive halos \citep{sheth:99, reed:03,  jenkins:01,
  warren:05}  and for an estimate of the covariance of \xray\
temperature and luminosity at fixed mass.    

For a set of structurally
identical, spherical halos with mass $\mdelta$, radius $\rdelta$ (here 
$\Delta$ is a threshold defining the halo edge relative to the
critical density) and with intracluster gas fraction $\ficm \se
\micm/\mdelta$, one expects the bolometric luminosity to scale as 
\citep{Arnaud:99} 
\begin{equation}
L_{\rm bol}(M,z) \ = \ C \, Q_L \, f^2_{\rm ICM} \ \rho^{7/6}_c(z) \ M_{\Delta}^{4/3} ,
\label{eq:LbolScaling}
\end{equation}
where $C$ is a constant, $\rhoc(z)$ is the critical density, and $Q_L \se (3/4
\pi) \int d^3y g^2(y)$ is a structure function that
is sensitive to the second
moment of the normalized gas density profile
\begin{equation} 
\rho_{\rm ICM}(r) =  \ficm \Delta \rhoc(z) \, g(r/\rdelta)  .
\label{eq:rhoICM}
\end{equation}

If the ensemble average gas fraction or radial structure function vary
with mass and epoch, then the power law exponents will generally
differ from the basic expectation.  This is the case
for a preheated ICM, with $L \spropto M^{11/6}$, invoked by
\cite{evrard:91} to reproduce the  X-ray luminosity function in an
$\Omega_m \se 1$ cold dark matter cosmology. Observations of the $L-T$ relation find a steeper slope, $L \propto T^3$, than expected from self-similarity 
\citep{mushotzky:97, fairley:00, novicki:02}.  Models incorporating 
angular momentum in halos have been able to reproduce the steeper 
relations \citep{delpopolo:05}.

In general, deviations from spherical symmetry and differences in
formation history and dynamical state will produce variations in
$\ficm$ and $g(y)$ among halos of fixed mass, leading naturally to some distribution 
$p(L|M,z)$ of soft band luminosity $L$.
We employ here a log-normal conditional likelihood, with
fixed dispersion $\sigl$ , centered on a power-law scaling relation 
\begin{equation}
L \ = \ L_{15,0} \  \rho_c^s(z) \  M^p .
\label{eq:LsoftScaling}
\end{equation}
It is important to note that this equation characterizes the log mean
behavior of a mass-limited population.  Surveys that are incomplete in
mass may differ from this. 

For a given choice of model parameters, cluster counts in 
flux and redshift are generated by convolving $p(L|M,z)$ with
the halo mass function of \cite{jenkins:01} and applying a mean fractional flux 
error correction.  Comparing to REFLEX survey  
counts \citep{reflex5} determines most likely parameters. 
Since the REFLEX survey is relatively shallow,
we choose not to solve for general redshift evolution.  Instead, we simply
compare cases of self-similar ($s \se 7/6$) and no ($s \se 0$)
evolution.   For soft X-ray luminosities, the expected mass and
redshift scalings are somewhat weaker than the bolometric case of
equation~(\ref{eq:LbolScaling}), $L \spropto M \rho_c(z)$, but we
employ $s \se 7/6$ as a slightly extreme variant.

Our analysis is therefore focused on the slope $p$, present-epoch
intercept, $L_{15,0}$, at $10^{15} \hinv \msol$, and the dispersion
$\sigl$, which we express in terms of the equivalent dispersion in
mass $\sigm \se \sigl/p$. 

A more direct approach to the problem of relating luminosity to mass
involves using individual cluster mass estimates derived under the
assumption of strict hydrostatic and/or virial equilibrium.
\cite{reiprich:02} (hereafter RB02) perform this 
analysis with 63 clusters from the flux-limited HIFLUGCS survey,
finding consistency with the power-law expectation,
equation~(\ref{eq:LsoftScaling}), with a slope slightly steeper than
self-similar and with a few tens of percent intrinsic scatter in mass
at fixed 
luminosity.   A promising future approach to this problem is to use
weak lensing mass estimates of stacked cluster samples
\citep{johnston:05}.   

In \S\ref{sec:model}, we explain the model and our approach to
analyzing the REFLEX survey counts.  In \S\ref{sec:results}, parameter
constraints that result from matching the counts are presented and
found to be strongly degenerate.  We
therefore apply a pair of additional constraints, one based on the
clustering of REFLEX clusters and the other on the observed scatter in
the temperature-luminosity relation. The latter exercise requires
input from cosmological simulations of clusters.    

We compare to results of RB02 and discuss implications for cluster
selection and hydrostatic mass estimates in \S \ref{sec:disc}. A summary and
conclusions are given in \S\ref{sec:conc}.  Except in \S\ref{sec:disc}, our
calculations assume a concordance 
$\Lambda$CDM  cosmology with $\Omega_m \se 0.3$, $\Omega_\Lambda \se
0.7$, ${\rm H}_0  \se 100 h \kmsmpc$, and $\sigma_8 \se 0.9$.

Throughout the paper, $L$ is the \xray\ luminosity in the rest-frame,
soft ROSAT passband ($0.1-2.4$ keV) in units of $10^{44}$ 
erg~s$^{-1}$ assuming $h \se 0.7$.  The mass $M$ is taken to be
$M_{200}$, the mass contained within a sphere whose mean interior
density is $200 \rhoc(z)$, expressed in units of $10^{15} \hinv
\msol$.   The reader is hereby notified that we employ different
Hubble constant conventions when presenting values for luminosity and
mass. 
  
\section{Modeling the Luminosity Function} \label{sec:model}
 
After briefly introducing the REFLEX cluster sample used in the
analysis, this section details our approach to modeling cluster counts 
in flux and redshift, including evolutionary corrections.

\subsection{The REFLEX Survey} \label{sec:reflex}

We use cluster counts from the REFLEX 
(ROSAT-ESO-Flux-Limited-X-ray) Cluster Survey, which covers 
an area of $4.24$ str and includes 
447 identified galaxy clusters down to a flux limit of 
3 $\times$ 10$^{-12}$ ergs 
s$^{-1}$ cm$^{-2}$ in the (0.1 to 2.4 keV) ROSAT band \citep{reflex5}.  
The \xray\ clusters have
been correlated with optical data from the COSMOS data base
with a relatively low threshold to define a highly complete cluster candidate
catalogue. Cluster validation and redshift determinations
were achieved by an intense spectroscopic follow-up
observing program at the European Southern Observatory, supplemented
by available literature. 

Luminosity determinations and the first REFLEX sample release 
are described in \cite{reflex1}.  Determination of the
survey selection function is described in \cite{reflex4} and
references therein.  We do not incorporate selection function
uncertainties directly into our analysis.  This simplification is
unlikely to significantly affect our conclusions, since at the high 
flux-limit of the REFLEX survey the sample is well understood.  
\cite{reflex1} estimate a loss under 5\% of clusters
in low exposure regions, and tests presented in that work (\cf
Figure~25) show that this changes  the derived luminosity function by
less than 2\%. 

\begin{figure}[t]
\centerline{\epsfysize3.5truein \epsffile{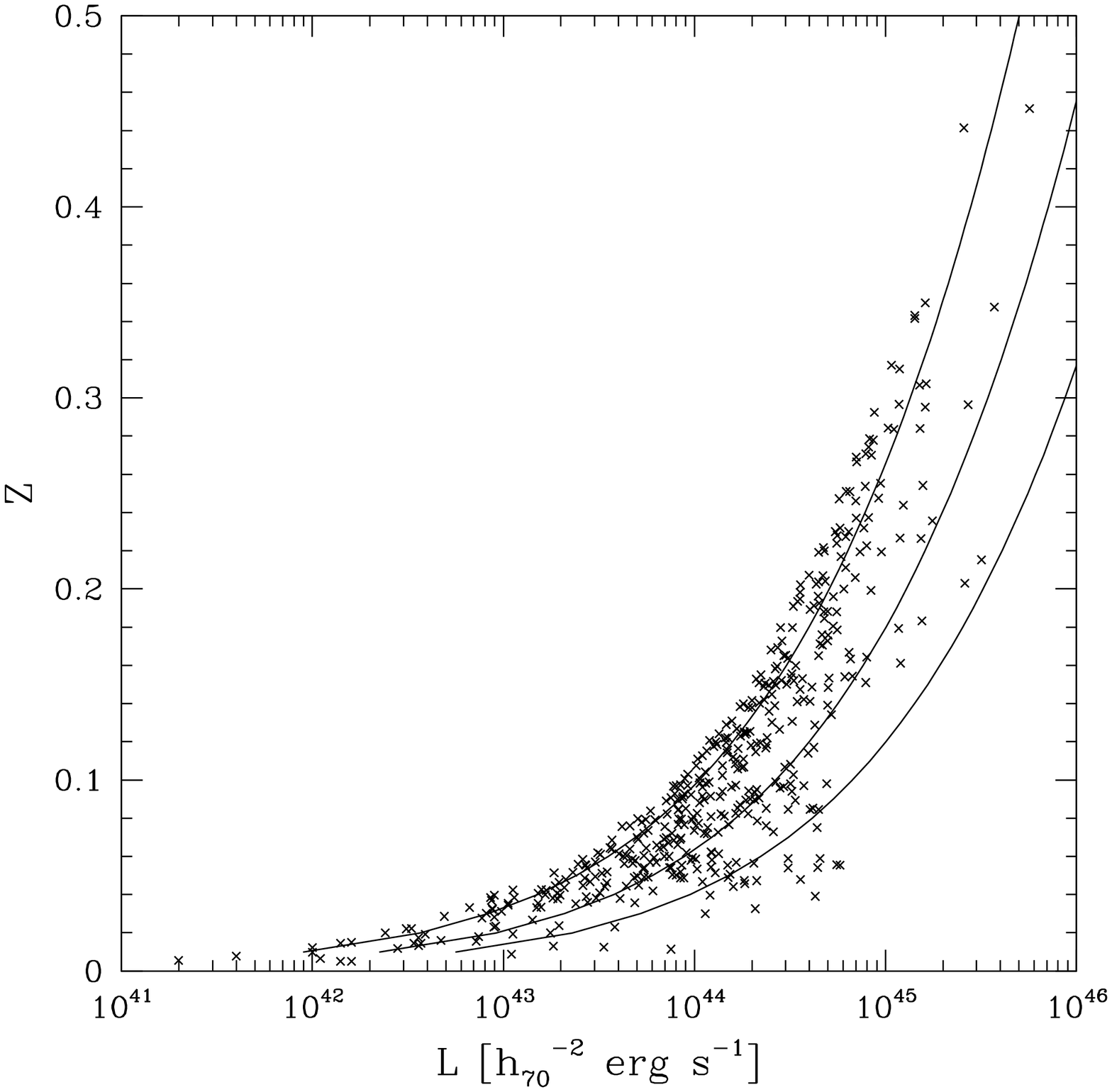}}
\caption{Redshifts and observed luminosities of 447 clusters in the
  full REFLEX survey.  The solid lines are lines of constant flux:
  4, 10, and 25 $\times$ 10$^{-12}$ ergs s$^{-1}$ cm$^{-2}$ in the
  observed (0.1 to 2.4 keV) ROSAT band.
\label{fig:logelz}}
\end{figure}

The redshifts and luminosities of the REFLEX clusters are 
shown in Figure~\ref{fig:logelz}.  The solid lines are lines of 
constant flux:  4, 10, and 25 $\times$ 10$^{-12}$ ergs 
s$^{-1}$ cm$^{-2}$ in the [0.1-2.4 keV] band. In the following
section, we perform a likelihood analysis on this count distribution
in flux and redshift.   

We must account for several survey effects when computing model expectations.  
Our basic model predicts the total flux while the REFLEX data 
are based on flux detected within the fixed angular aperture.   We
correct the data for this aperture effect by using the extrapolated
fluxes calculated by \cite{reflex1}.  This correction results in an
8\% average increase in luminosity, but for a few bright nearby
clusters the correction is larger.   

We also include a K-correction in our model.  The luminosity $L$ 
in our mass-luminosity relation is the emissivity integrated over the
rest-frame, 0.1-2.4 keV band, whereas the REFLEX data are derived from
the emissivity over a band redshifted by a factor $(1+z)$.  We find
that the K-correction used for the REFLEX survey construction, listed
in Table 3 of \citep{reflex5}, is well fit by the form  
\begin{equation}
K(z,T) \ =  \ \bigl( 1 + (1 + \log_{10}[T/5 \keV] \,z) \bigr)^{1/2}
\label{eq:Kcorr}
\end{equation}
where $L_{\rm obs} = K(z,T) \ L_{\rm rest}$.  We assign temperatures
to halos in a manner described below. 

\subsection{Theoretical Model}\label{sec:theo}

The space density of dark matter halos is now accurately calibrated
for the case of hierarchical clustering from an initially Gaussian
random density field.  We use parameters derived for the mass scale  
$M_{200}$ by \cite{evrard:02}, based on the Jenkins mass function (hereafter
JMF, \cite{jenkins:01}) form determined from a suite of cosmological
simulations.  

The JMF is described by a similarity variable 
$\sigma (M, z)$, where $\sigma^2$ is the variance in the linear
density field after top-hat filtering fluctuations on mass scale
$M$.  The function $\sigma(M,z)$ is a monotonic decreasing function of
mass for bottom-up models of structure formation and its inverse is
well fit in the log by a quadratic relation, $\ln
\sigma^{-1}(M,0) = s_0 + s_1 \lnM + s_2 (\lnM)^2$.  
Fit parameters for specific models we employ are listed in
Table~\ref{table:params}. \cite{jenkins:01} show that the mass
fraction in halos of mass $M$ at redshift $z$ 
\begin{equation}
f(M) \equiv \frac{M}{\rho_m(z)} \frac{{\rm d}n(M,z)}{\dln \sigma^{-1}},
\end{equation}
is well fit by 
\begin{equation}
f(M) = A \ \exp(-|\ln \sigma^{-1} + B|^{\epsilon}) .
\label{eq:JMF}
\end{equation}
When evaluating the mass function, we use the fact that $\sigma(M, z)$ scales
with the linear growth factor of density fluctuations $D(z)$.  In
addition, the parameters $A$, $B$ vary linearly with $\Omega_m$ as
described in \cite{evrard:02}.   The $z \se 0$ values are listed in
Table~\ref{table:params}.   

\begin{table}
\caption{Filtered Power Spectrum and JMF Parameters}
\label{table:params}
\begin{center}
    \leavevmode
    \begin{tabular}{lcccccc}
\hline \hline
{$\Omega_m$} & {$s_0$} & {$s_1$} & {$s_2$} & $A$ & $B$ & $\epsilon$ \\
\hline
0.24 & 0.468 & 0.267 & 0.0123 & 0.216 & 0.737 & 3.86 \\
0.30 & 0.549 & 0.281 & 0.0123 & 0.220 & 0.730 & 3.86 \\
0.36 & 0.608 & 0.287 & 0.0123 & 0.224 & 0.723 & 3.86 \\
\end{tabular}
\end{center}
\end{table}

In a small comoving volume element $dV$ at redshift $z$, we write the
probability $dp$ of finding a cluster of mass $M$ and luminosity $L$ as 
\begin{equation}
dp \ = \ p(L|M,z) \, n(M,z) {\dln}M {\dln}L \, dV
\label{eq:pjoint}
\end{equation}
We assume a log-normal conditional probability 
\begin{equation}
p(L|M,z) \ = \ \frac{1}{\sqrt{2\pi}\sigL} 
\exp\left[-\frac{(\lnL- \lnLbar)^2}{2\sigL^2} \right] 
\label{eq:Lpdf}
\end{equation}
with constant dispersion $\sigL$ and a log-mean (or median) luminosity that 
follows 
\begin{equation}
\lnLbar(M,z)  \ = \ {\ln}L_{15}(z) + p \, \lnM 
\label{eq:Lmed}
\end{equation}
where the normalization
\begin{equation} 
{\ln}L_{15}(z) = \lnlf + s\,{\ln}\left[\rhoc(z)/\rhoc(0)\right]
\end{equation}
sets the luminosity of a cluster
with mass $10^{15} h^{-1} M_\odot$ at redshift $z$. 

In principle, the
dispersion $\sigL$ could be variable in mass or redshift, but we
assume it to be constant, at least over the range of masses ($\gta
3 \times 10^{13} \hinv \msol$) and redshifts ($z \lta 0.3$) probed by the REFLEX
sample.  The same is true for the slope parameter $p$.  

The luminosity function at any epoch is the conditional probability
convolved with the JMF 
\begin{equation}
dn(L,z)  =  \int {\dln}M \ n(M,z) \, p(L|M,z) \, {\dln}L .
\label{eq:conv}
\end{equation}
The number counts of clusters at redshift $z$ with observed flux
greater than $f$ is  
\begin{equation}
\frac{dN(>f,z)}{dz} = \frac{dV}{dz}  \, \int_{L_{\rm min}(z)}^\infty dn(L,z) 
\label{eq:nfz}
\end{equation}
where $L_{\rm min}(z) \se 4\pi d_L^2(z)f / K(z,T)$, with $d_L(z)$ the
luminosity distance and $K(z,T)$ the K-correction fit of
equation~(\ref{eq:Kcorr}).  We use the analytic luminosity distance 
approximation presented in \cite{pen:99}. To calculate halo temperatures, we use
the relation of \cite{evrard:02} that reproduces the local temperature
function in a concordance model 
\begin{equation}
kT \  = \  6.5 \, \left(h(z)M\right)^{2/3} \, \sigma_8^{-5/3} \, {\rm keV} , 
\label{eq:TM}
\end{equation}
with $\h(z) \se h(0)  \bigl(\rhoc(z)/\rhoc(0)\bigr)^{1/2}$ the Hubble
parameter at redshift $z$.   For our choice of normalization, the
$z\se 0$ intercept at $10^{15} \hinv\msol$ is $7.7$~keV.   

With the log-normal assumption, equation~(\ref{eq:Lpdf}), the
flux-limited counts simply weight the mass function with a
complementary error function giving the fraction of halos that
satisfy the flux limit at each redshift
\begin{equation}
\frac{dN(>f,z)}{dz} = \frac{1}{2} \frac{dV}{dz}  \,  \int {\dln}M \
n(M,z) \,  \rm{erfc\left(y_{\rm min}(M,z)\right)} . 
\label{eq:nfzcum}
\end{equation}
with $y_{\rm min}(M,z) \se [{\ln}L_{\rm min}(z) - \lnLbar(M,z)]/\sqrt{2}\sigl$.

Our likelihood analysis is based on differential counts in flux and redshift 
\begin{equation}
\frac{d^2N(f,z)}{dz \, {\dln}f} \ = \ \frac{dV}{dz}   \int {\dln}M \ n(M,z) \,  p(L|M,z) 
\label{eq:nfzlike}
\end{equation}
with $L \se 4\pi d_L^2(z)f / K(z,T)$.   
 
\subsection{Flux Errors}\label{sec:fluxerror}

We must incorporate flux uncertainties in the REFLEX survey
into our analysis.  The mean fractional error in cluster flux 
is not large, $\langle \delta f/f \rangle \se 0.17$, but the individual errors
tend to increase at lower flux levels. 
Approximating the flux errors as constant, we modify
the theoretical predictions by convolving
the raw predicted counts with a Gaussian in ${\ln}f$ of fixed width 
$\sigma=0.17$.  This fractional scatter adds in quadrature to the
intrinsic model scatter.  

To avoid clusters with flux errors substantially larger than the mean, we
use a flux limit $4 \times 10^{-12} \cgsflux$ in 
the likelihood analysis, a few flux error sigma  higher than the formal 
REFLEX flux limit $3 \times 10^{-12} \cgsflux$.  This reduces the
sample size to 299 clusters but, because of the strong influence of the scatter
constraint discussed in \S\ref{sec:dlt} below, the uncertainties in our
final parameter estimates are not strongly affected by the reduced counts. 

As a check on this approach to incorporating flux errors, we have 
used an alternative that compares the
uncorrected theoretical counts to a Monte Carlo ensemble of REFLEX
realizations created using the set of measured flux errors.
The mean likelihood of the ensemble, using flux limit $4 \times
10^{-12} \cgsflux$, produces 
parameter constraints that lie within $1 \sigma$ of the results obtained using
the convolution  approach. We present our results using the latter
method.  

\subsection{Redshift Evolution}\label{sec:evoln}

Ideally, we would include the evolution parameter $s$ of
equation~(\ref{eq:LbolScaling}) as a free parameter, along with
$L_{15,0}$, $p$ and $\sigM$, in our analysis.   But because of 
the relatively shallow sample, and the lack of uniform luminosity and
redshift coverage (see Figure~\ref{fig:logelz}), we defer this
exercise to future, deeper samples.   

Instead, we limit our investigation to two distinct, and likely
extreme, cases: the self-similar (SS) model of
eq'n~(\ref{eq:LbolScaling}) and a no evolution (NE) model with
$L_{15}(z) \se L_{15,0}$.  Self-similar evolution is supported 
by recent work of \cite{maughan:05}, who investigate the redshift
behavior of the luminosity--temperature
relation in a sample of eleven clusters to redshift $z \sim 1$.
However, \cite{ettori:04} do not see 
strong evolution to $z \se 1.3$ in the luminosity--mass relation with a
sample of 28 clusters.

\subsection{Parameter Degeneracies}\label{sec:PLMF}

The convolution of the mass function is affected slightly differently
by the three model parameters. An increase in $L_{15,0}$ simply
shifts the luminosity scale to brighter values while the  
slope $p$ controls the local slope of
the luminosity function, with lower $p$ being steeper.  
Changing $\sigM$ produces a combination of the above
effects, affecting both the shape and normalization.  Over a finite range of
luminosity, changes induced by varying one parameter can thus be offset by
appropriate variations of the other two.  One therefore 
anticipates degeneracies among the model parameters when fitting to
counts alone.  

For the case of a pure power law mass function $n(M) \spropto
M^{-\gamma}$, we show in the Appendix that the degeneracy is exact and
takes the form 
\begin{equation}
C \ = \ \ln L_{15,0} + (\gamma p / 2) \sigM^2, 
\label{eq:powlaw}
\end{equation}
where $C$ is the quantity that is constrained by the data.

\section{Results}\label{sec:results}


After first presenting parameter constraints that result from matching
the REFLEX survey flux-limited counts, we extend the analysis to
include clustering bias and scatter in the luminosity-temperature
relation.  We then compare our model results to those of RB02, and
discuss the important role of flux-limited selection.   

\subsection{Constraints from REFLEX counts}\label{sec:mlm}

We calculate likelihoods based on the expected survey counts in narrow
flux and redshift bins, following the Poisson approach described in 
\cite{pier:01}.  We define confidence intervals 
relative to the best fit model using  
$\ln\left(\frac{\mathcal{L}}{\mathcal{L}_{\rm max}}\right) = - \chi^2/2$,
where $\mathcal{L}_{\rm max}$ is the maximum likelihood of all the models examined.

\begin{figure}[t]
\centerline{\epsfysize3.5truein \epsffile{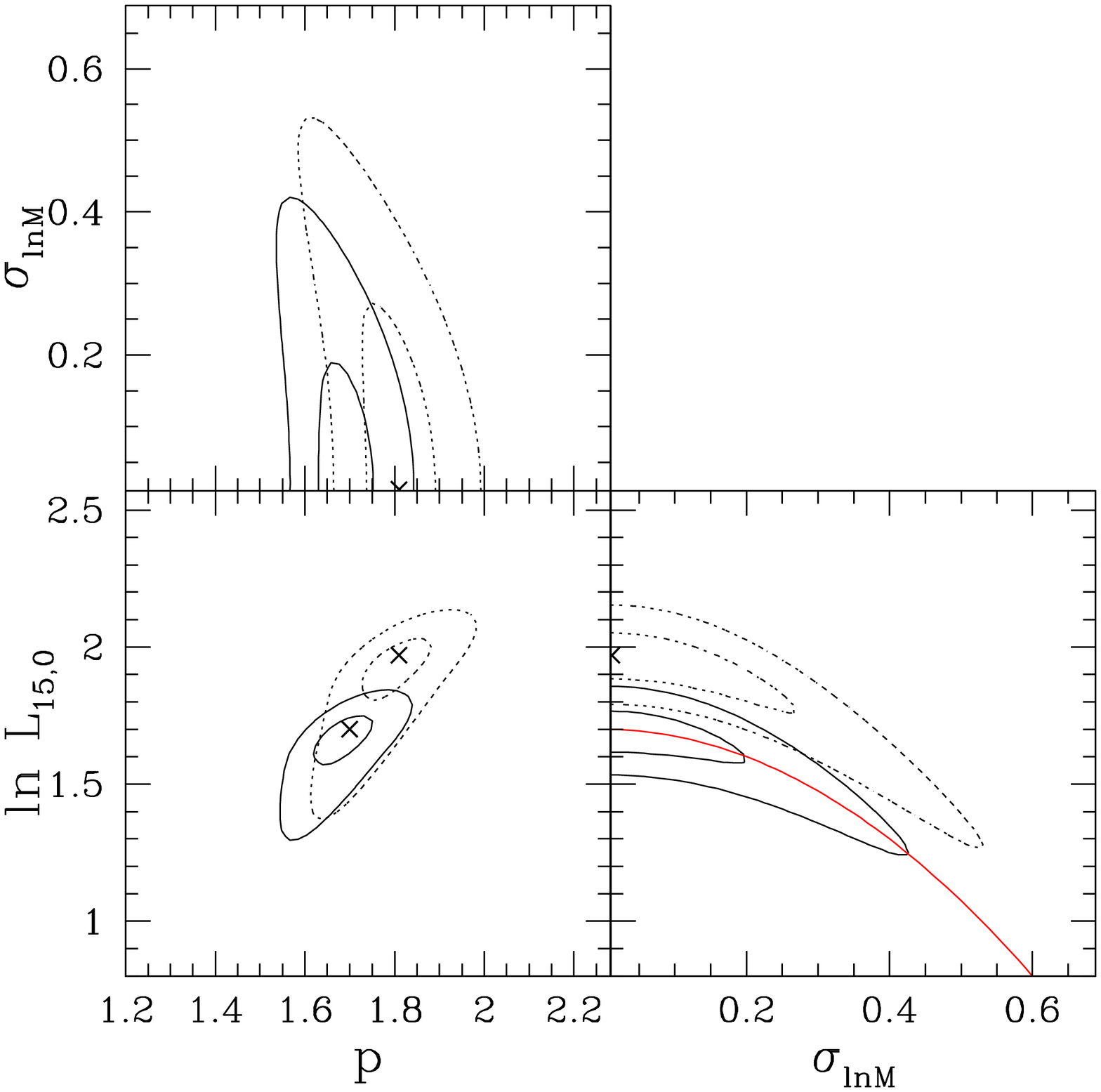}}
\caption{Contours of the 68 and 99\% marginalized likelihoods of the
  model parameters that result from matching the REFLEX counts
  are plotted for self-similar (solid) and no evolution
  (dashed) assumptions.   Crosses mark the maximum likelihood of the
  three-parameter model.  The added curve in the lower right-hand
  panel plots the degeneracy between $\dm$ and $\lnlf$ expected for a
  power-law number function, equation~(\ref{eq:powlaw}).
\label{fig:mlcounts}}
\end{figure}

Figure \ref{fig:mlcounts} shows 68 and 99\% confidence contours 
for the model parameters that result from matching the REFLEX survey
counts.  Each panel shows likelihoods marginalized over the missing
parameter.  Solid lines assume self-similar evolution of the luminosity 
while dashed lines assume no redshift evolution.  
Due to the degeneracy between the three parameters discussed
in section \ref{sec:PLMF}, the observed counts are reproduced at 99\%
confidence over a fairly broad swath of parameter space.  For the SS case, the slope can
take values $1.55 < p < 1.85$, intercepts lie in the range  $1.25 <
L_{15,0} < 1.85$, and the scatter is bounded by $\sigM  < 0.42$.
Zero scatter in the mass-luminosity relation, a seemingly 
unphysical solution given that cluster-sized halos are frequently  
dynamically active, is actually the most likely model, with slope $p \se
1.70$ and $\lnlf \se 1.70$. 

The NE model is shifted to a brighter intercept and steeper slope.  The
brighter intercept is needed to produce sufficient numbers of high
luminosity clusters at moderate redshifts, while the steeper slope
avoids overproducing counts at low luminosities and redshifts. 

\begin{figure}[t]
\centerline{\epsfysize 3.5truein \epsffile{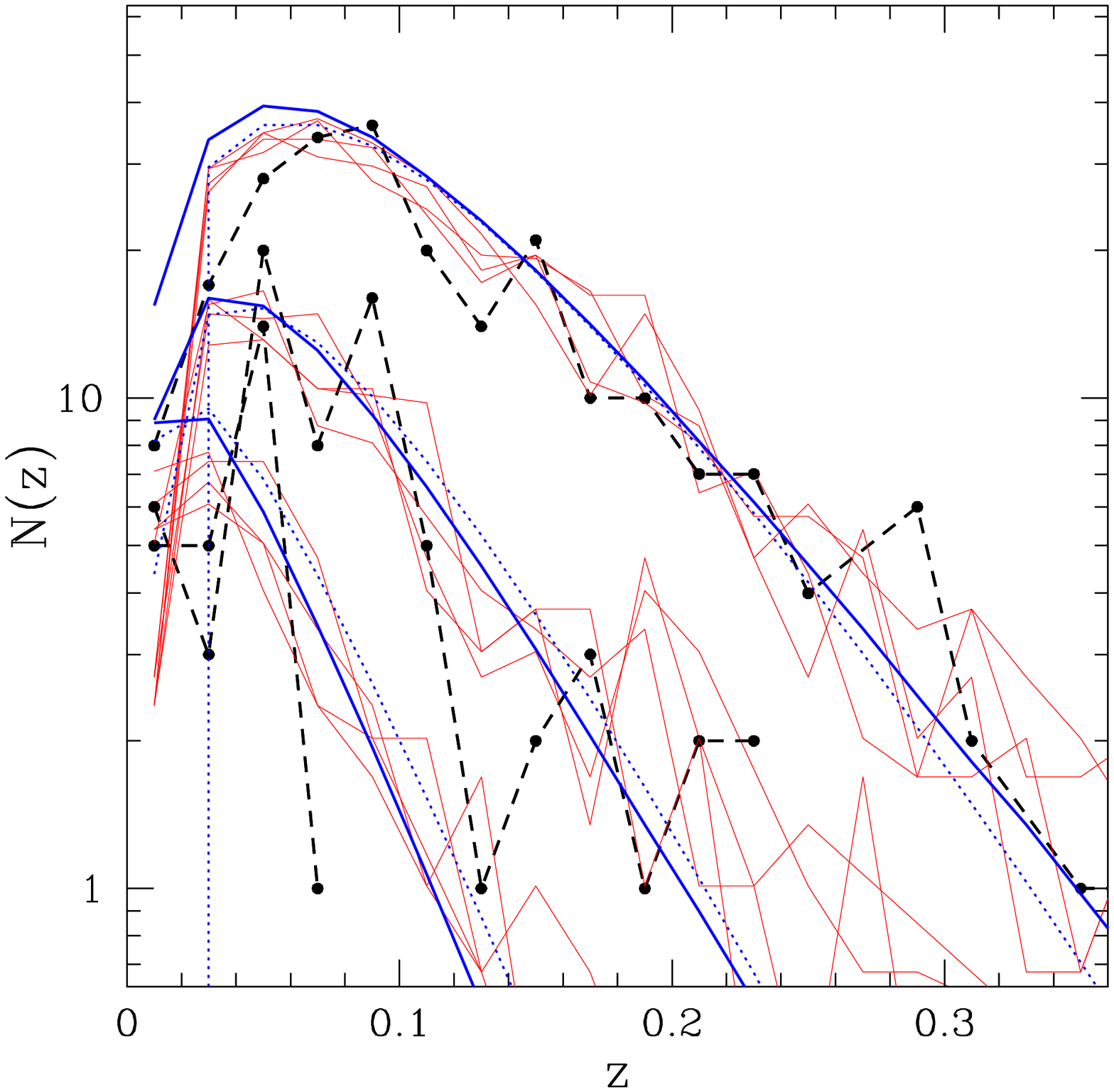}}
\caption{Differential counts in redshift for three flux ranges (from
  upper to lower):  $4 \le f < 10$, $10 \le f < 25$, and $f \ge  25
  \times 10^{-12} \cgsflux$.  The REFLEX data are plotted as filled
  circles, while the counts expected from the best-fit SS (solid) and
  NE (dotted) models are shown as continuous lines.   Light solid
  lines for each flux range show discrete realizations using discrete mock
  sky catalogs and the best-fit SS model.
\label{fig:nz}}
\end{figure}

Figure~\ref{fig:nz} shows that the best fit SS and NE models
provide good matches to the observed REFLEX counts in flux and redshift.
Also plotted are discrete counts derived from the Hubble 
Volume (HV) mock sky catalogs \citep{evrard:02}.  For each halo, we
assign a luminosity using parameters of the best-fit SS model, along
with a temperature using equation~(\ref{eq:TM}).   The K-corrected
flux is then calculated and used to determine counts in mock surveys with
REFLEX angular sky coverage.  The lines
in Figure~\ref{fig:nz} result from different random assignment of
luminosities to halos using the best fit final parameter discussed in
\S~\ref{sec:full}.

To improve these parameter constraints we must include additional
information from related observations or from 
theoretical priors derived, for example, from numerical simulations.  We
follow a largely empirical approach, and move on in the following
sections to add  
constraints from the effective bias of the REFLEX survey and from the 
scatter in the observed temperature--luminosity relation.
As explained below, interpreting the T-L scatter requires input
from cosmological simulations.

\subsection{Clustering Bias}\label{sec:bias}

Massive halos display stronger spatial clustering than the density field in
general (Kaiser 1984).  The halo--halo power spectrum $P_{\rm hh}(k)$
for mass-limited samples exhibits a nearly constant,  mass-dependent
bias $b(M) = \sqrt{P_{\rm hh}/P}$ on large scales.  We use the form
for this bias derived by \cite{sheth:99},  
\begin{equation}
b(M) = 1 + \frac{1}{\delta_c}\left[\frac{a_s \delta_c^2}{\sigma^2} 
- 1 + \frac{2 p_s}{1 + (a_s \delta_c^2 / \sigma^2)^{p_s}}\right]
\end{equation}
with parameter values $\delta_c = 1.68$, $a_s = 0.75$ and $p_s = 0.3$
from \cite{hu:03}.   The effective bias of a volume-limited sample
centered at (low) redshift $z$ is then
\begin{equation}
b_{eff} = \frac{\int \dln M \, w(M,z) \,n(M,z) \, b(M) }{\int \dln M \,
  w(M,z) \,n(M,z)} 
\end{equation}
where $w(M,z)$ is a weight function that gives the fraction of halos of
mass $M$ that lie above the minimum luminosity $L_{min}$ of the 
sample at redshift $z$.  We calculate the bias appropriate for the 
volume-limited sample L050 in \cite{reflex3}, with $L_{min} = 2.55
\times 10^{43} {\rm h_{70}^{-2}} erg \ s^{-1}$ and mean redshift $z
\se 0.04$.  We do not take into account any redshift dependence of the
bias, which will be small for this sample. 

\begin{figure}[t]
\centerline{\epsfysize 3.5truein \epsffile{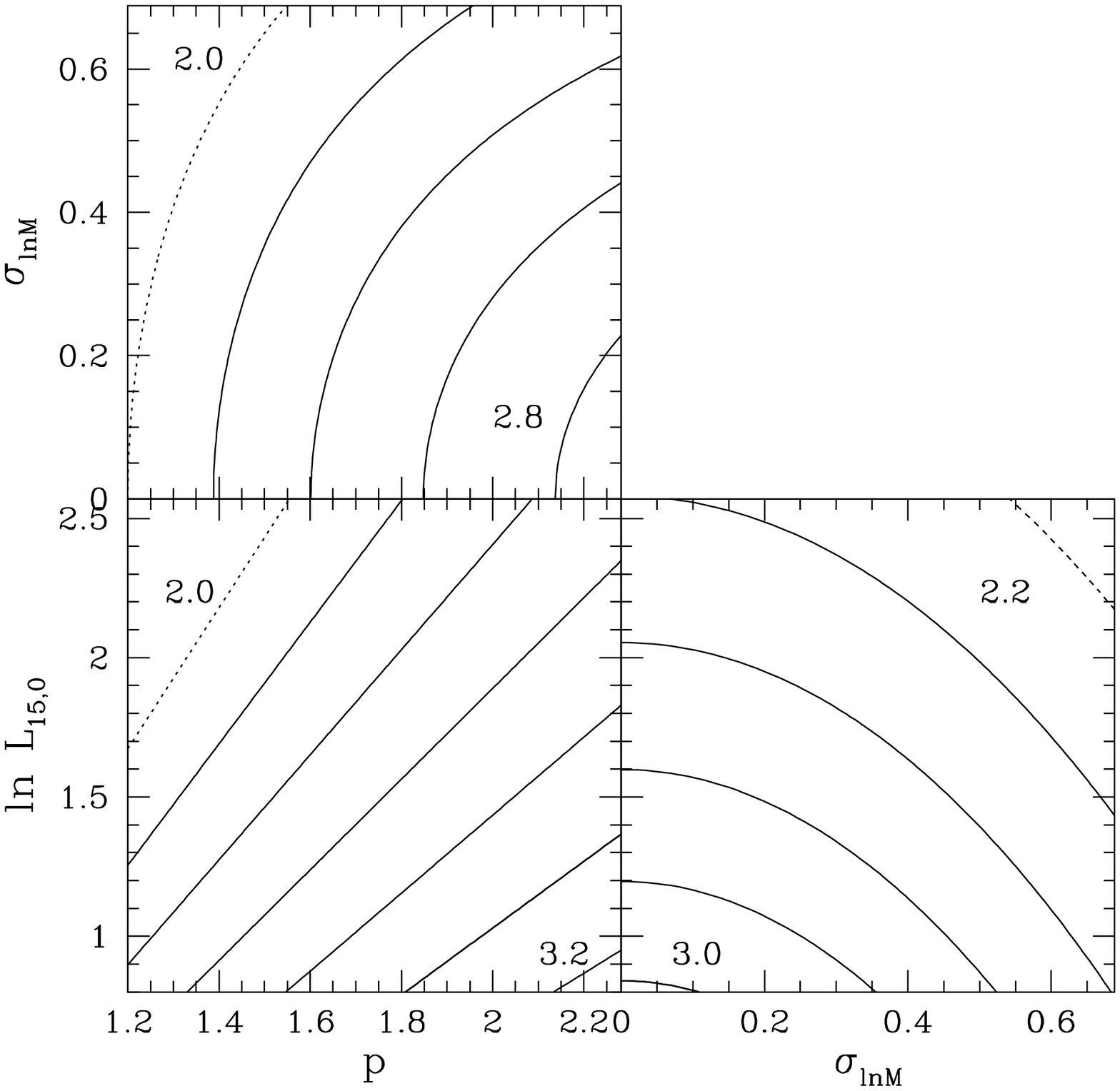}}
\caption{Contours of the effective bias $b_{eff}$ are shown in slices
through parameter space at the SS best-fit location of
Fig~\ref{fig:mlcounts}.  Contours are spaced by $0.2$.
 \label{fig:bias}}
\end{figure}

Figure \ref{fig:bias} plots contours of $b_{eff}$ for slices within
the model parameter space centered on the best-fit location of the SS
model in Figure~\ref{fig:mlcounts}.  As $p$ increases, the range of
halo masses sampled decreases, implying a higher minimum mass scale
and therefore an increasing bias.  The effective bias decreases as
$\ln L_{15,0}$ increases due to the fact that lower mass halos satisfy
the flux limit.  Finally, increasing \dm causes a decline in the
effective bias, because increased scatter adds  lower-mass halos at a
fixed luminosity. 

The power spectrum and uncertainties published for this sample by
\cite{reflex3} imply a prior on the effective bias of $b_{eff} =
3.0 \pm 1.0$ for a concordance cosmology.  
As shown in Figure \ref{fig:bias}, 
the $1\sigma$ uncertainty on the effective bias nearly fills the 
parameter space we are considering.  Because of the large 
uncertainty, adding the bias constraint 
does not significantly improve the fits from counts alone.  
More precise measurement of the clustering bias will require 
deeper surveys that better probe the high mass/high bias portion of
mass function.

\subsection{Scatter in the Luminosity-Temperature Relation}\label{sec:dlt}

\begin{figure}[t]
\centerline{\epsfysize 3.5truein \epsffile{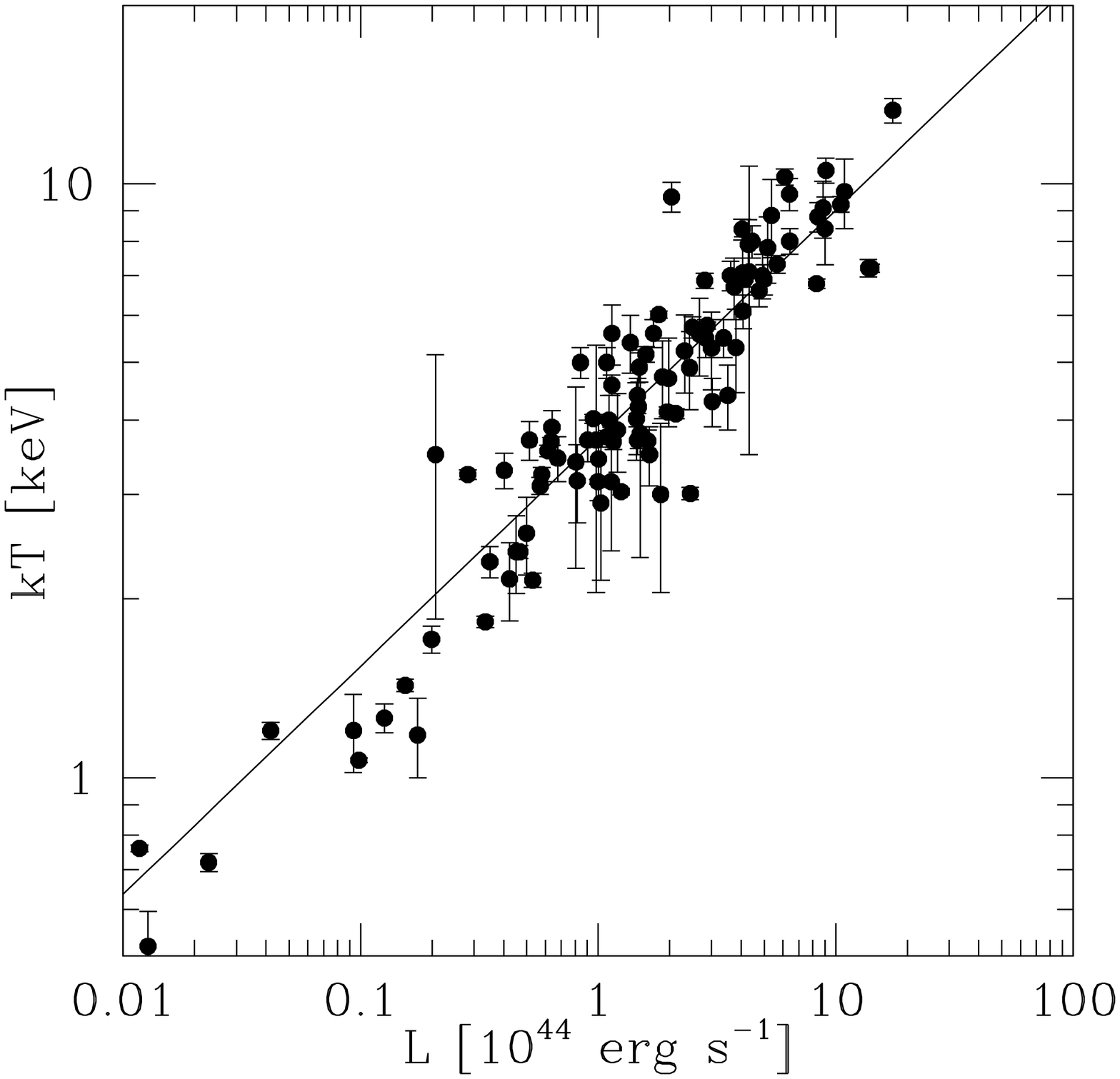}}
\caption{The T-L relation for HIFLUGCS clusters from
\cite{reiprich:02}.  Error bars indicate the observed $1\sigma$
uncertainties in the temperature measurements.
\label{fig:lxt}}
\end{figure}

We next derive information on the mass variance $\sigma_{{\ln}M}^2$ from the
intrinsic scatter of the temperature--luminosity relation.  
Figure~\ref{fig:lxt} shows the T-L relation for
106 HIFLUGCS clusters published by RB02, with temperatures obtained
from various original sources.  We fit the relation to a power law,
including the stated errors in temperature, and then measure the
intrinsic variance $\varTL$ by subtracting the measurement
error contribution from the total variance.  We bootstrap over the
quoted errors on the temperatures to estimate the uncertainty in the
intrinsic dispersion, with the result $\sigTL = 0.25 \pm 0.03$.
While there are other cluster surveys with luminosity and temperature
data \citep{hornerphd:01, ikebe:02}, we use temperatures from RB02, to 
be consistent throughtout the paper.

To include the measured T-L scatter in the analysis, consider a
mass-limited ensemble of co-eval halos with masses, \xray\ spectral
temperatures and soft \xray\ luminosities $\{M_i,T_i,L_i\}$.
Motivated by the virial theorem, and mirroring the case for
luminosity, we assume that $T$ and $M$ are related by a power-law with
log-normal scatter, and write $\overline{\lnT} \se {\ln}T_{15} + q
\lnM$.  Any given halo will, in general, deviate from the mean L-M
and T-M scalings.  The $i^{\rm th}$ cluster's true mass ${\ln}M$ will
deviate from the values inferred by inverting the log-mean relations
for its temperature and luminosity 
\begin{eqnarray} 
&\ln M &= \frac{1}{q} \ln \left(\frac{T}{T_{15}}\right) + \delMT,  \\
&\ln M &= \frac{1}{p} \ln \left(\frac{L}{L_{15}}\right) + \delML.
\end{eqnarray}

The halos temperature and luminosity are then related by
\begin{equation}
{\ln}T \ = \ \frac{q}{p}{\ln}L + {\ln}T_{15} - \frac{q}{p}{\ln}L_{15} + q(\delML - \delMT) ,
\end{equation}
so, considering a subset of halos of fixed luminosity $L$,  the
expected T-L relation is  
\begin{equation}
\langle \lnT\rangle \ = \  \frac{q}{p}  \ \lnL + C , 
\label{eq:TL}
\end{equation}
where $\langle \, \rangle$ represents an average at fixed luminosity.  The intercept 
$C  \se {\ln}T_{15} - \frac{q}{p}{\ln}L_{15} + q[ \langle \delML
\rangle -  \langle \delMT \rangle]$ will be biased from the
zero-scatter expectation if the mean mass deviations based on
temperature and luminosity do not cancel, $\langle \delMT \rangle \ne
\langle \delML \rangle$.   

The second moment $ \langle ( {\ln}T -  \langle \lnT\rangle )^2
\rangle$ of the observed T-L relation yields  
\begin{equation}
\varTL  = q^2 \bigl[  \langle (\delMLhat)^2 \rangle  - 2 \langle
\delMLhat \, \delMThat \rangle +  \langle (\delMThat)^2 \rangle \bigr], 
\label{eq:covar}
\end{equation}
where  $\delMLhat \sequiv  \delML - \langle \delML \rangle $ and
$\delMThat \sequiv  \delMT - \langle \delMT \rangle $ are deviations
from the sample means, again at fixed luminosity.    

We show below that flux-limited samples are biased in favor of
low-mass halos $\langle \delML \rangle \! < \! 0$, but we also find
that the measured mass variance in mock HIFLUGCS samples is nearly
unbiased relative to a complete, mass-limited sample.  We therefore
assume  $\varML \se \langle \delMLhat^2 \rangle$.  
Noting that the variance of our underlying model is $\varM \se
\varML$, then equation~(\ref{eq:covar}) can be rearranged to give 
\begin{equation}
\varM = \frac{\varTL}{q^2} - \varMT + 2 \langle \delML \, \delMT \rangle .  
\label{eq:varM}
\end{equation}

This equation links the variance in mass at fixed luminosity to an
observable quantity, the T-L variance $\varTL$, along with other
terms that are not directly observable.   To evaluate these terms, we
turn to an ensemble of 68 cluster simulations evolved under a
`preheated' assumption \citep{bialek:06}.  Like previous simulations
\citep{evrard:96, bryan:98, math:01, borgani:04, rasia:04}, these
models respect the virial scaling between temperature and mass,
displaying a logarithmic 
mean relation $T \spropto M^{0.58 \pm 0.02}$ that is slightly
shallower in slope than the self-similar value of $2/3$ due to the
enhanced effects of the raised initial entropy on low mass systems.
To approximate actual temperature estimates, the temperature $T$ is 
determined from fitting mock \xray\ spectra of each simulated cluster,
assuming emission from a plasma enriched to 0.3 solar metallicity.

\begin{figure}[t]
\centerline{\epsfysize 3.5truein \epsffile{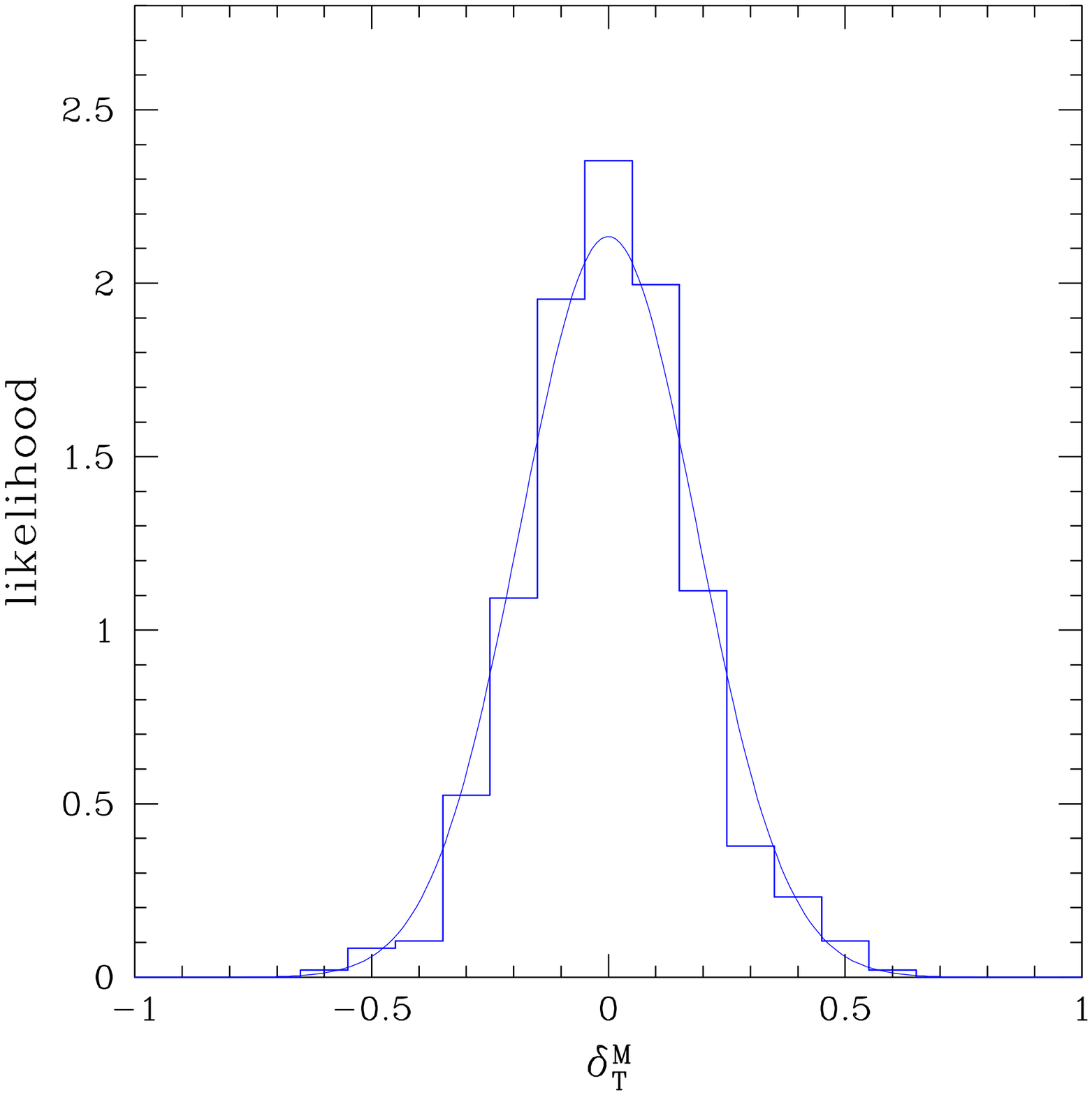}}
 \vskip-0.1in
\figcaption{ The histogram shows deviations in (natural) log mass from
  mean power-law
  fits to the $M-T$ relation from 68 preheated cluster
  simulations \citep{bialek:06}.  The solid line shows a Gaussian
  fit with zero mean and dispersion of $0.19$ \label{fig:MdevHist}}
\end{figure}

Figure~\ref{fig:MdevHist} shows the frequency distribution of the
residuals in mass estimates based on inverting the mean virial
relation for the 68 clusters sampled at six low-redshift 
epochs, $z \se 0.290$, $0.222$, $0.160$, $0.102$, $0.49$ and 0. The
distribution of logarithmic deviations is very close to Gaussian, with 
dispersion $\sigMT \se 0.19$.  This small scatter is empirically
supported by the work of \cite{mohr:99}, who find $\sims 17\%$
scatter in $M_{ICM}$ at fixed temperature.   

\begin{figure}[t]
\centerline{\epsfysize 3.5truein \epsffile{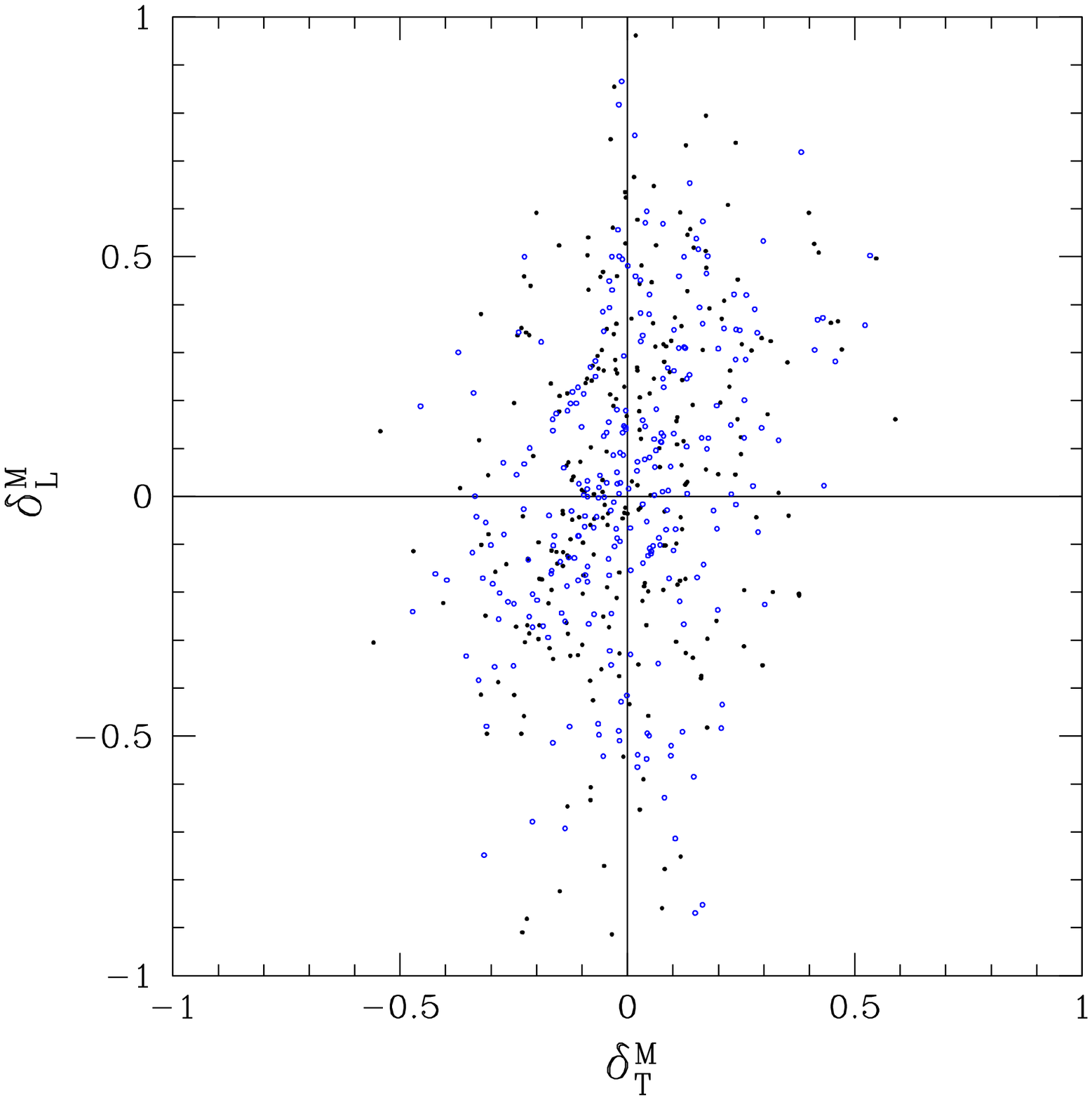}}
 \vskip-0.1in
\figcaption{ The correlation of residuals in log mass about
  the mean M-T and M-L relations derived from 68 preheated cluster
  simulations \citep{bialek:06} at epochs $0<z<0.2$ (filled
  circles) and $0.2<z<0.3$ (open circles). \label{fig:Mresid}}
\end{figure}

We use data from the same models to examine the covariance in the M-L and M--T 
deviations, shown in Figure~\ref{fig:Mresid}.  The simulations display weak covariance
$ \langle \delMT \, \delML \rangle = 0.017$.  The weak coupling 
presumably arises because $T$ and $L$ are sensitive to 
processes operating on different scales and because the effects of
multiple mergers, rather than a single dominant encounter, are driving
the halo dynamics at most times.   This finding may seem at odds with
results from binary merger simulations by \cite{ricker:01}.  In the binary
case, the luminosity and temperature rise and fall in concert as the 
mass increases during the merger, suggesting strong covariance.  
However, recent cosmological simulations by \cite{rowley:04} show
complex evolution in $L$ and $T$ during cluster evolution that is more
aligned with the results of Figure~\ref{fig:Mresid}. 

With these estimates in hand, and with $\sigTL$ 
from the RB02 data, we find the scatter in mass 
at fixed luminosity from equation~(\ref{eq:varM}) to be
\begin{equation}
\sigM \ = \ 0.43 \pm 0.06 , 
\label{eq:scatVal}
\end{equation}
where we have assumed a T-M slope $q \se 0.58 \pm 0.05$,
equivalent to $M\spropto T^{1.72 \pm 0.12}$.  
The slope uncertainty, along with the
intrinsic T-L scatter measurement error, dominate 
the uncertainty in $\sigM$.  Note that a self-similar T-M scaling, 
$q \se 2/3$, leads to a $1\sigma$ reduction in scatter, $\sigM \se 0.37$.

\subsection{Full Constraint Results}\label{sec:full}

\begin{figure}[t]
\centerline{\epsfysize 3.5truein \epsffile{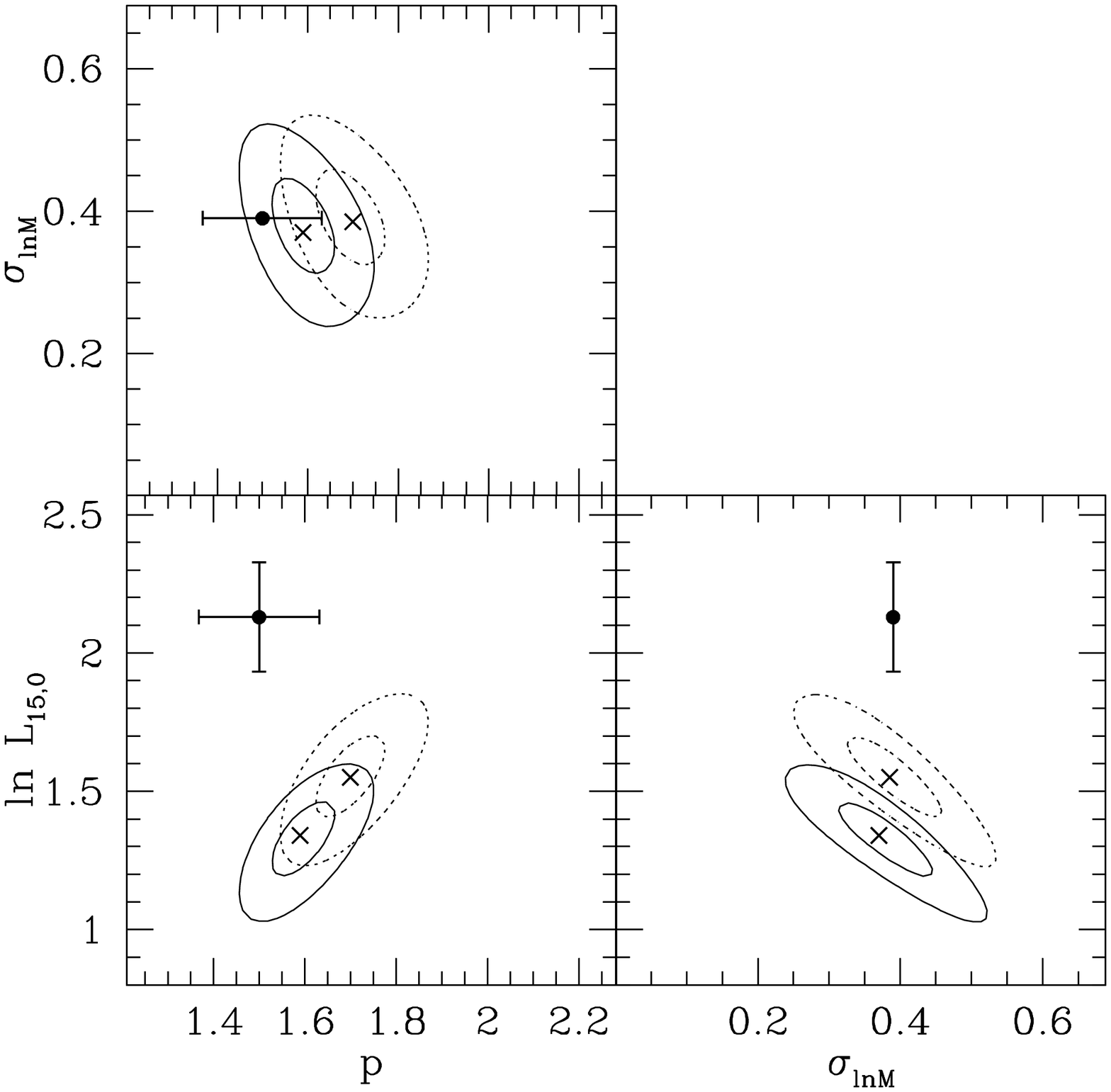}}
\caption{Marginalized 68 and 99\% confidence intervals are shown
after including constraints from clustering bias and from the
estimated scatter of equation~(\ref{eq:varM}).  Solid and dotted
lines are SS and NE models.  Crosses mark the maximum likelihood
location.  The data point is the RB02 result, shown with 90\%
confidence error bars on the slope and intercept.
\label{fig:mlmpribs}}
\end{figure}

\begin{figure}[t]
\vspace{-2.0 truecm}
\centerline{\epsfysize 3.5truein \epsffile{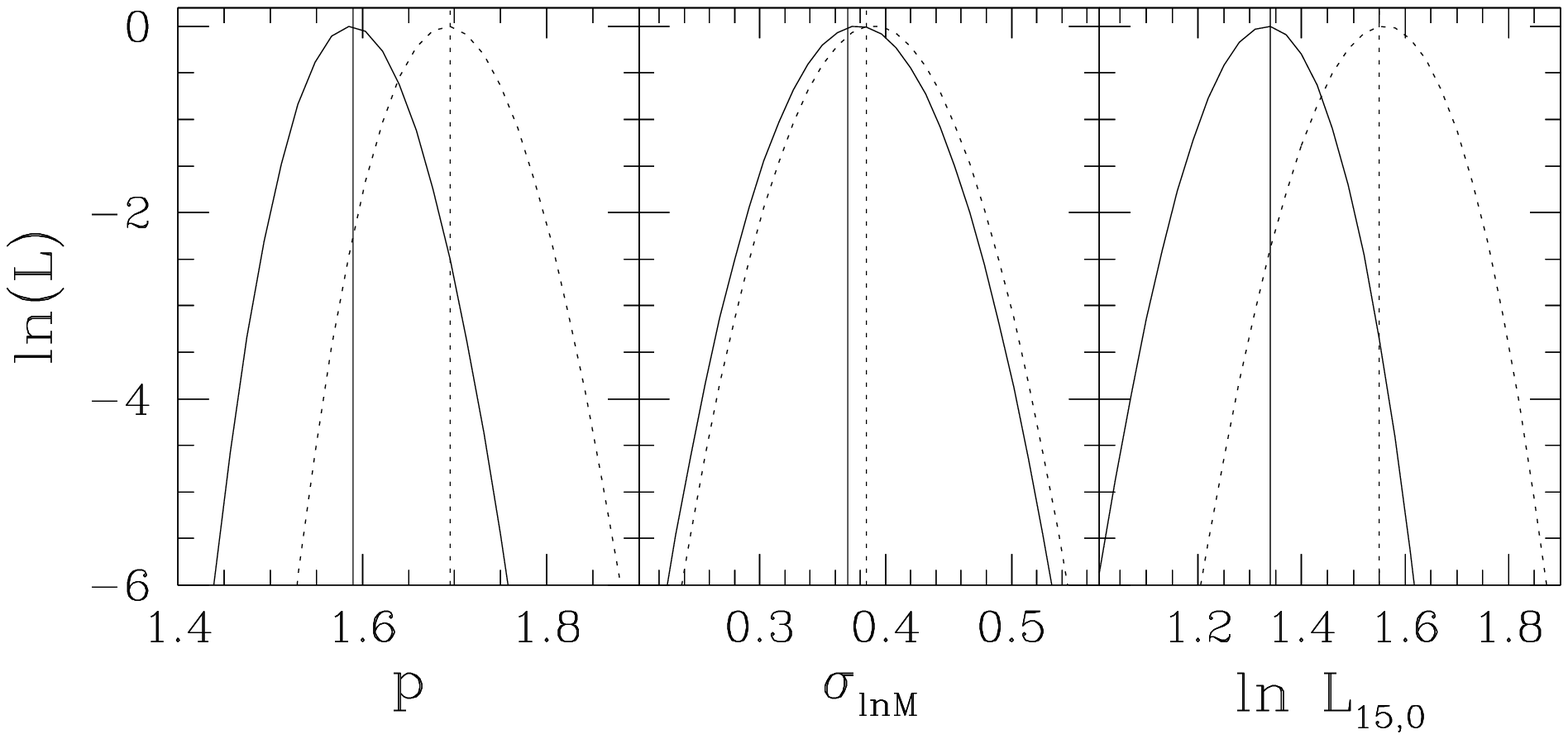}}
\vspace{-2.0 truecm}
\caption{One-dimensional, marginalized likelihoods for each model parameter
for the SS (solid) and NE (dotted) models after applying the
constraints used in Figure~\ref{fig:mlmpribs}.  Vertical lines denote
the maximum likelihood location in the full parameter space.
\label{fig:marg2d}}
\end{figure}

Adding the constraint of equation~(\ref{eq:scatVal}), along with the
effective bias, to the  
analysis produces the marginalized likelihoods plotted in 
Figures~\ref{fig:mlmpribs} and \ref{fig:marg2d}.   This analysis yields
values $p \se 1.59 \pm 0.05$, $\lnlf \se 1.34 \pm 0.09$, and
$\sigM  \se 0.37 \pm 0.05$.  These values are strongly dependent on  
the scatter constraint, and thus depend on the measurements of temperature, 
luminosity, and their quoted errors, all of which determine $\sigTL$. 

The solid circle in Figure~\ref{fig:mlmpribs} shows parameter values from RB02.  
Because the published values assumed $\Omega_m \se 1$, we have made
small (typically few percent) corrections to convert their luminosity and mass 
estimates to a concordance cosmology.  Under an isothermal profile
assumption ($\rho(r) \spropto r^{-2}$) used by RB02,   
the correction to cluster mass scales as
$\rho_c^{-1/2}(z)$, while the luminosity scales as $d_L^2(z)$, with
$d_L$ the luminosity distance.  

Fitting corrected luminosity to mass, we find best-fit values $p \se 1.50 \pm
0.08$ and $\lnlf \se 2.13 \pm 
0.12$ for the slope and intercept of the RB02 sample.  Subtracting in
quadrature the measured  
mass uncertainties from the scatter in the L-M relation, we find
an estimate of the intrinsic scatter, $\sigM = 0.39$.  
A formal uncertainty in this estimate is complicated by the fact
that there are several sources of scatter, including variance in the
hydrostatic mass estimator, that are not easily separated.  We defer
this exercise to future work. The fact that this estimate of the
intrinsic M-L scatter agrees with our model results suggests that the
scatter in hydrostatic mass estimates is less than $\sigM$.    

The RB02 normalization result is in serious conflict with the value
determined here. 
We address the main source of this discrepancy, and discuss other
implications, in the following section.   

Note that the slope, $q/p$, inferred for the T-L relation
via equation~(\ref{eq:TL}) is $0.36 \pm 0.03$, in reasonable agreement
with the $T \sim L^{0.45 \pm 0.04}$ displayed by the data in
Figure~\ref{fig:lxt}.   A self-similar slope, $q \se 2/3$, yields
better agreement, $q/p \se 0.42$, but the data do not compel this choice.

\begin{figure}[t]
\centerline{\epsfysize 3.5truein \epsffile{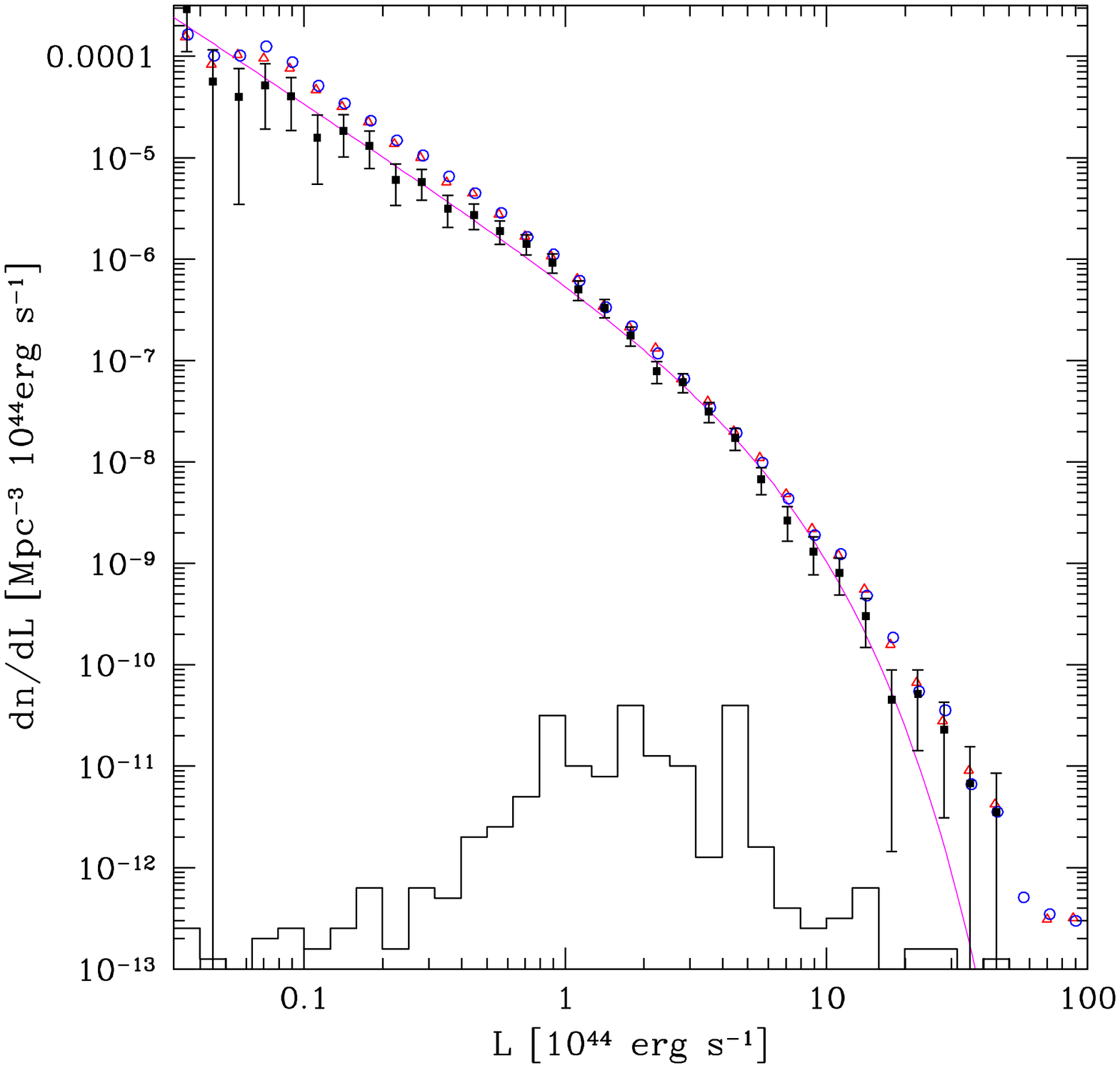}}
\caption{The luminosity function of the REFLEX data (filled squares),
  shown with $1\sigma$ uncertainty from Poisson and flux errors, is
  compared to mock HV sample results using the best fit SS (open
  circles) and NE (open triangles) parameters from
  Figure~\ref{fig:mlmpribs}.  The solid line is the Schecter function
  fit to the luminosity function from \cite{reflex1}.  The histogram
  shows raw REFLEX counts in 0.1 dex luminosity bin, using a linear
  scale of 10 counts per decade.
\label{fig:nlfull}}
\end{figure}

As a further comparison of the model and data, we show in 
Figure \ref{fig:nlfull} a binned representation of the luminosity function of 
the REFLEX data, determined with a $1/V_{\rm max}$ sum described in
\cite{reflex1}, along with discrete realizations of the function
derived from HV mock sky catalogs using the best-fit SS and NE
parameters (circles and triangles, respectively).  Errors in the
binned data include contributions from flux errors and Poisson
fluctuations.  At low luminosities, the sample variance due to the
small volume probed becomes significant.   The histogram shows raw
REFLEX counts in 0.1 dex luminosity bins. 

The mock realizations provide a good match to the REFLEX
luminosity function above $\sims 4 \times 10^{43} \ergs$, where the
survey counts per bin exceed ten.  At lower luminosities, 
the models predict higher space densities than those observed. The
small numbers of clusters makes it difficult to
assess the nature of these differences.  It may signal a
breakdown in our model, perhaps through a steepening of the slope $p$
at low luminosities.  Alternatively, 
the REFLEX survey may be slightly incomplete at the low
luminosity end, which is populated mainly by nearby, extended systems
near the survey flux limit.  This interpretation is supported by the
finding of a `local void' within $z \lta 0.03$ in the cluster
spatial distribution \citep{reflex3}.
At the bright end,
both the observed and model counts 
decline more slowly than a pure Schechter form, $dn/dL \spropto
e^{-L/L_\ast}$.

\section{Discussion}\label{sec:disc}

In this section, we first discuss the source of the discrepancy
between the M-L relations derived here and that of RB02.  We then
briefly explore sensitivity of the model parameters to cosmology,
focusing on changes to $\sigate$ and $\Omega_m$.  We then 
comment on theoretical uncertainty in the mass function,
discuss connections to previous REFLEX analysis, and finish with
remarks on the accuracy of isothermal beta mass estimates.  

\subsection{Selection Bias from Flux Limit}\label{sec:flim}

The results presented in Figure~\ref{fig:mlmpribs} indicate that the
independent estimate of the luminosity--mass relation from RB02, based on 
isothermal $\beta$-model binding mass estimates, disagrees with the
best-fit L-M relation derived above.  There is reasonable agreement
in the slope and scatter, but the RB02 intercept is substantially
brighter than the best-fit values of both the SS and NE model
results. 

The large discrepancy in \lnlf\ results largely from a
classical Malmquist bias expected in flux limited samples such as
HIFLUCGS.  For a given mass and redshift, a portion of the tail of
low-luminosity clusters is lost due to the application of the flux
limit.   The amplitude of the resultant bias depends on the scatter in
the mass-luminosity relation.  We first calculate the expected bias
analytically, then demonstrate it using mock realizations of
flux-limited samples.   

With our model, the expected log mean
luminosity ($y \sequiv \lnL$) as a function of mass ($x \sequiv lnM$)
for a flux-limited sample is 

\begin{eqnarray}
  \langle y(x) \rangle & = 
 \frac{1}{dN_f(x)} \int dV(z) \, n(x,z) \, Y(x,z) \\
 Y(x,z) & = \frac{1}{2} \bar{y}(x,z) {\rm
     erfc}(t_{\rm min}) + \sqrt{\frac{1}{2\pi}}  \sigma_y e^{-t_{\rm
       min}^2} 
\label{eq:lnLbar}
\end{eqnarray}
where $dV(z)$ is the comoving volume element at redshift $z$, $t_{\rm
  min} \se (y_{\rm min}-\bar{y}(x,z))/\sqrt{2}\sigma_y$, 
$y_{\rm min} \se {\ln}L_{\rm min}(x,z)$, $\bar{y}(x,z) \se
\lnLbar(x,z)$, and $\sigma_y \se \sigl \se p \, \sigM$.  As before,
$L_{\rm min}(x,z) \se 4\pi d_L^2(z)f / K(z,T)$ is the K-corrected
minimum luminosity required to satisfy the flux limit at redshift
$z$. The normalization $dN_f(x)$ is the differential number of halos
of mass $e^x$ that satifsy the flux limit
\begin{equation}
  dN_f(x) \ = \ \frac{1}{2} \int dV(z) \, n(x,z) \, {\rm erfc}(t_{\rm min}) .
\label{eq:Nx}
\end{equation}

\begin{figure}[t]
\centerline{\epsfysize 3.5truein \epsffile{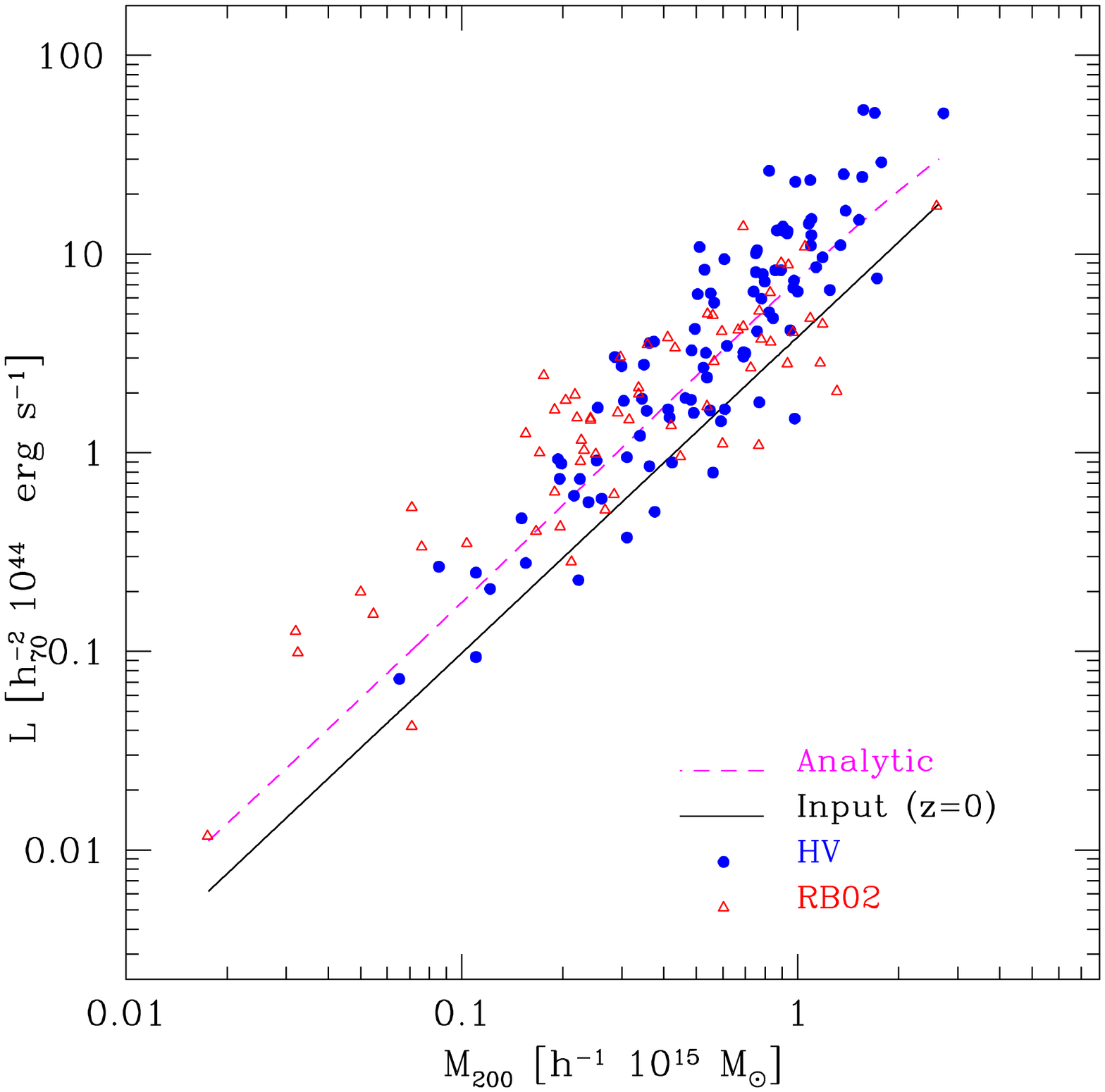}}
\caption{The best fit SS model relation between rest-frame, soft X-ray
  luminosity and mass at $z \se 0$ (solid line) is compared to the
  relation expected from a HIFLUCGS flux-limited sample (dashed line),
  computed from equation~(\ref{eq:lnLbar}).  Triangles show the
  HIFLUGCS data while filled circles show an HV mock version of these
  data.
\label{fig:LMflim}}
\end{figure}

We employ this model, with the JMF assuming a concordance cosmology
with our best-fit SS parameters, to calculate the L-M relation
expected for the HIFLUGCS survey, with ROSAT soft-band flux limit
$2 \times {\rm 10^{-11} erg s^{-1} cm^{-2}}$.   The result is shown
as the dashed line in Figure~\ref{fig:LMflim}.   The HIFLUGCS sample
  systematically selects brighter objects at fixed mass, leading to a
  brightening, by roughly a factor two, in the L-M relation relative
  to the present epoch relation, shown as the solid line.   
 
\begin{figure}[t]
\centerline{\epsfysize 3.5truein \epsffile{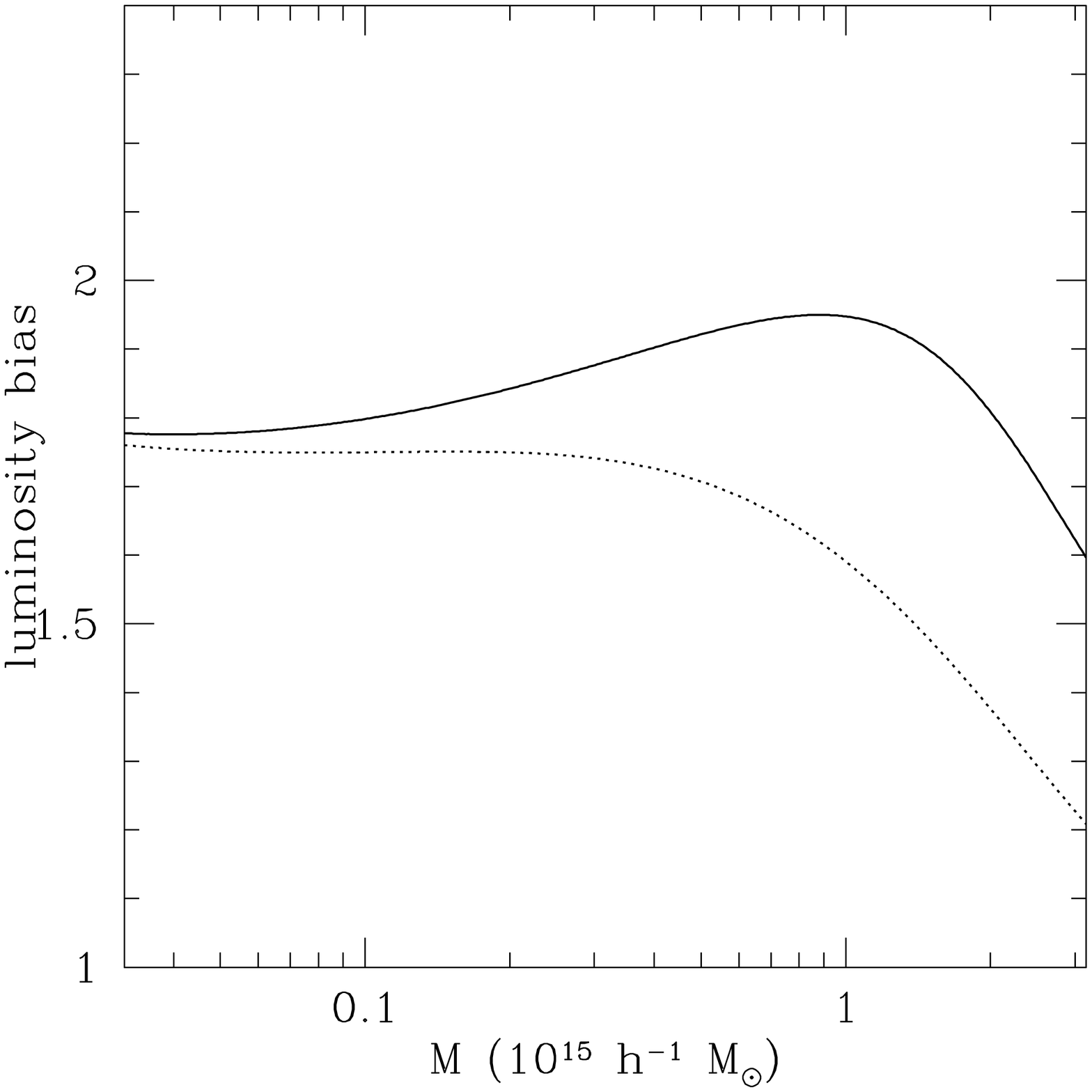}}
\caption{The ratio between the logarithmic mean luminosity of a $1.7 \times {\rm
    10^{-11} erg \ s^{-1} cm^{-2}}$ flux-limited sample and the $z \se
  0$ relation is shown for the best-fit SS (solid) and NE (dotted)
  models.  The difference in the two models is the result of redshift
  evolution, which is more important for higher mass halos that
  satisfy the sample flux limit at larger distances.
\label{fig:LMbias}}
\end{figure}

The brightening for this high flux sample results mostly from scatter,
and not from evolution in the SS model.  Figure~\ref{fig:LMbias}
demonstrates this by comparing the increase, relative to $z\se 0$, in
log mean luminosity, 
calculated from equation~(\ref{eq:lnLbar}) for SS and NE models.   At
low masses and luminosities, the volume 
probed is small, and there is little evolution in the SS model.  Above
$10^{14} \hinv \msol$, the survey depth is approaching $z \ssim 0.1$,
and the SS model bias becomes larger than that of the NE model.  As
the survey probes to higher redshifts at higher masses, the 
SS model bias continues to grow up to $M \ssim 10^{15} \hinv \msol$.
Beyond this mass scale, the 
bias drops because the number of the most massive halos falls
to zero at high redshift.   A similar decline is seen in the NE model.   

In addition to the analytic calculation, we also create mock
realizations of the HIFLUGCS sample using HV sky survey samples.  We
assign a luminosity and temperature to each halo in a manner
consistent with our best-fit model, and apply the flux cut to
K-corrected fluxes.  The result for one random realization is shown 
as filled circles in Figure~\ref{fig:LMflim}.   The points scatter
about the analytic expectation with {\sl rms} deviation of $\sigL 
\se 0.63$, close to the input (mass-limited) scatter of
$0.59$.   Due to the finite size of the HIFLUGCS sample, the degree
of bias will vary from the analytic expectation.  We estimate this
variance using 200 Monte Carlo realizations of HV mock samples,
derived from the MS and VS sky surveys \citep{evrard:02},
adjusted to a sky area of 5.5 str to match the cluster counts in 
RB02.  From this exercise, we derive correction terms of $-0.70 \pm
0.11$ to $\lnlf$ and $-0.06 \pm 0.07$ to $p$.   The correction terms 
strongly depend on the degree of scatter in the L-M relation.

\subsection{Degeneracy and Cosmology}\label{sec:cosm}

The normalization $\lnlf$, amount of scatter $\sigM$, and power spectrum
normalization $\sigate$ are strongly coupled
parameters that control the number counts.  In the current  
approach, we determine the scatter via measurement of variance in the
L-T relation, using input from simulations in a manner described in
\S\ref{sec:dlt}.  The rather large scatter in luminosity for halos of
a given mass, $\sigL \se 0.59$, implies the signficant Malmquist
bias discussed above.  Previous analysis of the REFLEX cluster 
abundance and power spectrum \citep{reflex3, 
  schuecker:03} assumed a much smaller level scatter, $\sigL \ssim
0.2-0.3$, based on the impression that the main contribution
to the L-M scatter derived in RB02 was large uncertainty in
the binding mass measurement.  With the smaller assumed scatter,  
no Malmquist bias correction was applied, so the value of $\lnlf$
employed was higher, and the corresponding value of $\sigate$ lower,
compared to the values used here.  This approach has recently been
emphasized by \citep{reiprich:06}.  

The set of parameters allowed by Figure~\ref{fig:mlmpribs} assumes a
concordance cosmological model described in \S~\ref{sec:intro}.  
We now explore the sensitivity of the best-fit location to changes in
cosmological parameters $(\Omega_m, \sigate)$, under a flat-metric
assumption.  In particular, we focus below on the 
world model favored by year-three WMAP analysis \citep{spergel:06},
which involves a lower normalization $\sigate \se 0.76 \pm 0.05$ and matter
density $\Omega_m \se 0.234 \pm 0.035$.  

\begin{figure}[t]
\centerline{\epsfysize 3.5truein \epsffile{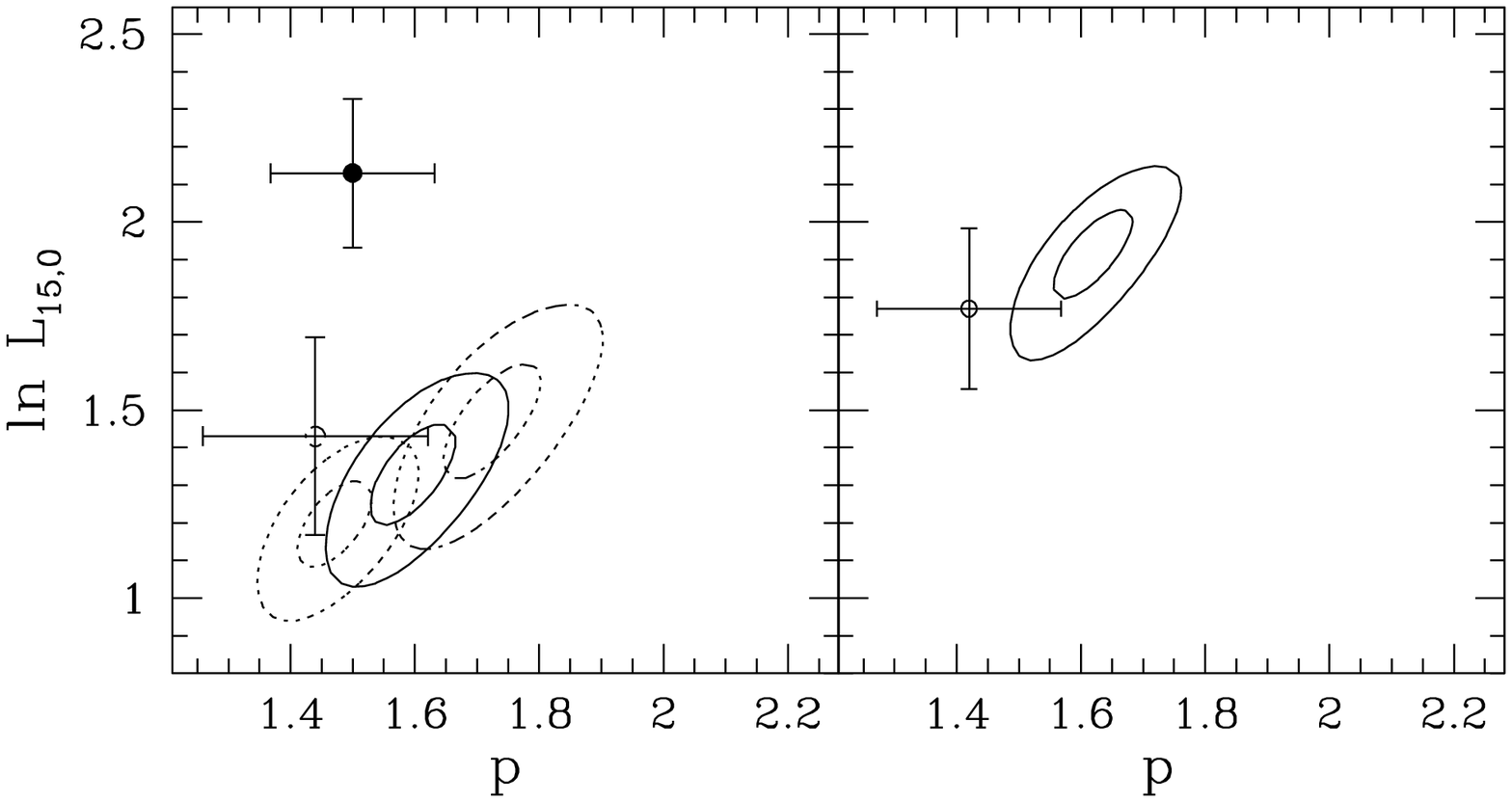}}
 \vskip-0.1in
\figcaption{The left panel plots 68 and 99\% confidence intervals
of the slope and normalization for $\Omega_m = 0.24$,
$0.3$ and $0.36$ (left to right), with $\Omega_m \sigate^2$ constant,
after applying the joint
constraints on counts, clustering, and $\sigM$ scatter. The solid point in
the left panel plots the original
RB02 result, with 90\% error bars.   The
right panel plots 68 and 99\% confidence intervals of the slope and normalization
for the WMAP compromise cosmology, $\Omega_m = 0.24$, $\sigate =
0.85$, after applying the
joint constraints on counts, clustering, and a lower $\sigM \se 0.25$
scatter.  In both panels, the open point gives the RB02 result
after correcting for the flux cut bias, with 90\% error bars that
include the variance of HV Monte Carlo realizations discussed in the text.
\label{fig:varomps}}
\end{figure}

The left panel of Figure~\ref{fig:varomps} shows how the results of our 
likelihood analysis change as the cosmology is varied.  
The dotted and dashed lines show 68 and 99\%
marginalized contours for $\Omega_m = 0.24$ and $\Omega_m 
= 0.36$, respectively.   The power spectrum normalization is varied to
keep $\sigma_8^2 \Omega_m$ constant, a condition that roughly holds
fixed the number of halos above $\sims 2 \times 10^{14} \hinv\msol$ 
(Efstathiou, Frenk \& White 1993; Viana \& Lyddle 1998).  The vacuum
energy density 
$\Omega_\Lambda \se 1-\Omega_m$ is adjusted to maintain a flat metric. 
The shape of the filtered power spectrum $\sigma (M)$ is determined
from CMBFAST \citep{cmbfast} and we use equation~(11) of 
\cite{evrard:02} to adjust the constants $A(\Omega_m)$ and 
$B(\Omega_m)$ in the JMF.  Parameter values are listed in 
Table \ref{table:params}.

The constraint from the observed scatter in the T-L relation ties
the best-fit location to a narrow range of $\sigM$.  However, as
$\Omega_m$ increases, the slope of the mass function steepens, and the
best fit value of $p$ must increase as a result.  We find that the
slope scales roughly as $p \spropto \Omega_m^{0.5}$, being reduced to
$p = 1.46 \pm 0.04$ for $\Omega_m = 0.24$ and increased to $1.71 \pm
0.05$ for $\Omega_m = 0.36$ for the case of SS evolution.
Table~\ref{table:fitpars} lists the best fit SS parameters for these
cosmologies.   

\begin{table}
\caption{SS Model Parameters for Different Cosmologies}
\label{table:fitpars}
\begin{center}
    \leavevmode
    \begin{tabular}{lccc}
\hline \hline
{$\Omega_m$} & {$\lnlf$} & {$p$} & {$\sigM$} \\
\hline
0.24 & $1.19 \pm 0.08$ & $1.46 \pm 0.04$ & $0.39 \pm 0.05$ \\
0.30 & $1.34 \pm 0.09$ & $1.59 \pm 0.05$ & $0.37 \pm 0.05$ \\
0.36 & $1.46 \pm 0.11$ & $1.71 \pm 0.05$ & $0.37 \pm 0.05$ \\
\end{tabular}
\end{center}
\end{table}

For the most massive and \xray\ luminous halos, a change of
normalization $\sigate$ produces a  
nearly constant shift in the mass function in the mass direction, such
that the mass $M_\ast$ at fixed space density $n_\ast$ scales as
$M_\ast \spropto \sigma_8^{1/\alpha^\prime}$, with $\alpha^\prime \se
0.4$ above $\sims 2 \times 10^{14} \hinv\msol$ \citep{evrard:02}.    
Since observations fix the luminosity at $n_\ast$, variations in
$\sigate$, and hence the mass scale $M_\ast$, can be compensated by
adjusting the normalization $L_{15,0} \spropto M_\ast^{-p}$.  We
therefore expect $L_{15,0} \spropto \sigma_8^{-p/\alpha'}$.  With
$\alpha' \ssimeq 0.4$, and $p \simeq 1.6$, the characteristic
luminosity should have strong sensitivity to the power spectrum
normalization, $L_{15,0} \propto \sigma_8^{-4}$.  

The open symbol in the left panel of Figure~\ref{fig:varomps} shows the RB02
result, after correction for the flux bias described in the preceding
section.  Error bars have been enlarged by adding in
quadrature the scatter in the corrections derived from the mock HV
Monte Carlo analysis.  The corrected results lie very close to the
values determined here.   

The right hand panel offers a compromise model driven by cosmological
parameters from the WMAP Year 3 data \citep{spergel:06}.  The model
assumes a flat metric with $\Omega_m = 0.24$ and a slightly elevated 
normalization $\sigma_8 = 0.85$, along with a reduced estimate of the
scatter $\sigma_{\ln M} = 0.25 \pm 0.06$.  The resulting best fit
parameters are $p = 1.60 \pm 0.05$, $\sigma_{\ln M} = 0.21 \pm 0.06$, 
and $\lnlf = 1.92 \pm 0.08$.  The lower scatter leads to a smaller
Malmquist bias for the RB02 sample, we derive corrections of $-0.36
\pm 0.05$ to $\lnlf$ and  
$-0.08 \pm 0.04$ to $p$.  The corrected RB02 result is then $p = 1.42 \pm 0.09$, 
$\lnlf = 1.77 \pm 0.13$, and is plotted in the right-hand panel in
Figure ~\ref{fig:varomps}.  

We concur with the analysis by \cite{reiprich:06} that a zero-scatter
solution with $\sigma_8 = 0.76$ provides good agreement with the
original RB02 result, but we feel that a zero-scatter solution is
unphysical.  The large variance in the observed T-L
relation has been tied to variations in core emission
\citep{fabian:06, ohara:06}.  Maintaining a tight L-M relation would
require a strong coupling between core gas physics and total halo
mass, as well as a
large scatter in the virial (T-M) relation.  These outcomes are neither 
anticipated theoretically nor supported by the current generation of
numerical simulations.   

\subsection{Theoretical Uncertainty}\label{sec:thunc}

Our model rests on the calibration of the $M_{200}$ halo space density
\citep{evrard:02}.  A full treatment of the theoretical error in the
mass function is beyond the scope of this paper, but we comment here
on the impact of normalization uncertainties in $n(M,z)$ on the
$\lnlf$ parameter.  

Simulations by \cite{huKravtsov:03} find agreement with the form we
employ at better than ten percent in number over the mass 
range $5\times 10^{13}-10^{15} \hinv\msol$ (see their Fig.~B10).
The JMF calibration using percolation, rather than spherical
overdensity, masses has been
confirmed at the five percent level in number by \cite{warren:05} using a
set of sixteen $1024^3$-particle simulations.  That work proposes a slightly
more complex formulation to replace our use of
equation~(\ref{eq:JMF}), but the difference at high masses is only a few
percent in number (see their Fig.~2).  

These results appear to indicate that the current calibration
uncertainty in the halo space density is at the level of $\sims
5\%$ in number.  For massive, \xray\ luminous halos, where locally
$d{\ln}n/d\lnM \ssimeq -3$, this 
translates into an error in mass scale of $\sims 2\%$, or an
additional  $\sims 0.03$ uncertainty in $\lnlf$.  

However, \cite{warren:05} also find that one out of four
non-concordance simulations shows a larger discrepancy with the
model expectations, roughly $20\%$ in number.  Considering that this
is one outlier out of 20 simulations, and given that the remaining
$N$-body data show much better consistency, we feel that a $15\%$
uncertainty in number density is a conservative upper limit on the
theoretical error.  This implies a maximum  
additional uncertainty in $\lnlf$ of $\sims 0.08$.  The theoretical 
uncertainty is most likely smaller than, or, at worst, comparable
to, the statistical error of $0.09$.  

The use of $M_{500}$ as a mass measure is not expected to affect the 
theoretical uncertainties.  Assuming an NFW mass profile, Appendix B 
of \cite{evrard:02} calculates a simple shift to the mass scale.  
Shifting the mass scale results in a shift  of the JMF, and our 
normalization parameter would likewise be translated.

\subsection{Accuracy of Hydrostatic Mass Estimates}\label{sec:massest}

The isothermal $\beta$-model masses employed by RB02 are
determined by fitting the azimuthally-averaged \xray\ surface
brightness to derive the gas density profile under the assumption of
isothermality for the intracluster gas.  Simulations indicate that
this approach may yield masses that are biased by up to a few
tens of percent \citep{evrard:90, schindler:96, nfw:95, evrard:96, 
rasia:04, rasia:06}, but the computational expectations are rendered uncertain  
by incomplete modeling of ICM physics and by systematic errors
associated with determining the spectral temperature derived from the
thermally complex intracluster medium of simulated
clusters \citep{mathiesen:01, mazzotta:03, vikhlinin:05}.  The 
uncomfortably wide range of inferred mass--temperature relations from
theory and observations (\cf Table~1 of \cite{pierpaoli:03}, Table~3
of \cite{henry:04} and Table~1 of 
\cite{evrard:04}) is another reason for caution regarding the 
accuracy of cluster mass estimates.   

From Figure~\ref{fig:varomps}, it is apparent that implied errors
in hydrostatic masses will be model-dependent.  For the
concordance model with $\sigate=0.9$, there is little error in normalization at $10^{15}
\hinv \msol$ after taking the flux bias into account.  For the WMAP 
cosmology, we tuned agreement in the right panel of
Figure~\ref{fig:varomps} by choosing $\sigate \se 0.85$ and $\sigm \se
0.25$.  
For a power spectrum normalization of $\sigate \se 0.76$, our normalization 
$\lnlf$ is brighter than the RB02 result by a factor of $1.5$, indicating that hydrostatic masses 
are overestimates of the true cluster mass by nearly a factor of 2. 

Additional independent estimates of the slope and intercept of the L-M
relation would help.  Weak lensing analysis of joint \xray\ and
optical samples is a promising approach that has been applied in a
preliminary fashion to clusters in the SDSS Early Data
Release \citep{sheldon:04}.  The
lensing mass estimates require careful calibration to eliminate the
effects of projection \citep{metzler:01}, but either stacking to
acheive spherical symmetry \citep{johnston:05} or 
analysis of mock data from suitably selected simulated clusters
\citep{hennawi:05} can be
used to validate procedures used to extract mass estimates from shear
maps.   \cite{pedersen:06} examine the M-T relation at the high-mass end 
using lensing masses 
from 30 clusters, finding a higher normalization than studies using 
isothermal masses, and finding a power spectrum normalization of 
$\sigma_8 \sim 0.88$ for $\Omega_m = 0.3$.  
The large number statistics offered by nearby optical cluster catalogs
from SDSS \citep{bahcall:03, miller:05, koester:06} and 2dF 
\citep{padilla:04}, when matched with existing and forthcoming
archival \xray\ data, should ultimately be exploited for this exercise.

\section{Conclusion}\label{sec:conc}

We derive constraints on a model that relates soft \xray\ luminosity
to halo mass and epoch,  
$p(L|M,z)$, described as a median power-law relation, $L \spropto M^p
\rho_c^s(z)$, with log-normal scatter of fixed width $\sigL$.
Convolving this with the mass function of a $\Lambda$CDM universe, 
we perform a likelihood analysis on the counts of galaxy
clusters in flux and redshift observed in the REFLEX survey for
specific cases of self-similar ($s \se 7/6$) and no evolution ($s=0$).
A strong degeneracy between the intercept and scatter in the model
means that counts place only an upper limit on the degree of scatter
in mass at fixed luminosity $\sigM  < 0.4$.  

To improve the estimate of the scatter, we apply additional
constraints based on the clustering bias of a volume-limited REFLEX
sub-sample and on the intrinsic variance of the observed 
temperature--luminosity relation.  Because the survey probes halo mass
scales where the bias is relatively weakly dependent on mass, and
because the measured clustering has a substantial uncertainty, the addition of bias has
little practical effect on allowed parameters.   Using gas dynamical
simulations of clusters to probe the inter-relationships of ${M,T,L}$,
we derive an estimate $\sigM  \se 0.43 \pm 0.06$ from the observed
scatter in the T-L relation.   

After applying these additional constraints, we derive best-fit
parameters $p = 1.59 \pm 0.05$, $\sigM  = 0.37 \pm 0.05$, and $\lnlf =
1.34 \pm 0.09$ for a concordance cosmology with $\{\Omega_m,
\Omega_\Lambda, \sigate\} \se \{0.3,0.7,0.9\}$ and a self-similar
evolutionary assumption.  The slope and normalization increase, by
$2\sigma$ and $1\sigma$ respectively, under a no evolution assumption.  
Exploring sensitivity to the assumed (flat-metric) cosmology, we find
$\elf \ssim \sigate^{-4}$ for fixed $\Omega_m$ and $p \sims
\Omega_m^{0.5}$ when $\Omega_m \sigate^2$ is held fixed.  When applying a cosmological model guided by WMAP results, $\Omega_m = 0.24$, $\sigate = 0.85$, and a lower constraint 
on the scatter, we derive best-fit parameters which include a 
much higher normalization term, $p = 1.60 \pm 0.05$, 
$\sigM = 0.21 \pm 0.06$, $\lnlf = 1.92 \pm 0.08$.  

Our concordance-model result differs from that determined by
RB02 using hydrostatic (isothermal-beta model) mass
estimates, being lower in normalization by roughly a factor of two.
We show that a bias of this magnitude is expected for the direct L-M
relation derived from a flux limited sample, due to a Malquist bias
induced by the survey selection process.  Correcting for this bias
leads to quite good agreement between the two approaches, implying
that systematic uncertainties in hydrostatic mass estimates are
unlikely to be larger than $\ssim 15\%$.  To accommodate the change in
cosmology preferred by WMAP observations, we offer a compromise
solution with slightly elevated $\sigma_8$ and lower intrinsic scatter
$\sigm$.  
Independent approaches to determining the L-M scaling relation,
particularly through weak gravitational lensing analysis, should be vigorously
pursued.    

We anticipate that the approach used here can be fruitfully applied to
relations between other observables, such as ${\rm X-ray}$ temperature or
Sunyaev-Zel'dovich decrement, and underlying mass.  The coming era of
large area, deep cluster surveys, coupled with increasingly detailed
numerical simulations, will drive progress, for example, by allowing
the evolutionary parameter $s$ to be directly constrained and by
enabling investigation of the detailed form of the likelihood
$p(L|M,z)$.  Possible future refinements for the latter include adding
a non-Gaussian component and/or relaxing the assumptions of constant
slope and scatter.  

\bigskip
We acknowledge Kerby Shedden for input regarding 
the observed flux errors.  This work was supported by NASA grant
NAG5-13378 and by NSF ITR  
grant ACI-0121671.  Support for the Virtual Cluster Exploratory,
from which the simulation data were
obtained, is acknowledged from  {\it Chandra} Theory grants
TM3-4009X and TM4-5008X.  AEE acknowledges support from the Miller
Foundation for Basic Research in Science at University of California,
Berkeley.  

\clearpage

\appendix
\section{Appendix}  

Consider the case of a power-law mass function at some epoch 
\begin{equation}
n(M) \, d\lnM \ = \ n_{15} \ M^{-\gamma} \, d\lnM,
\end{equation}
where $M$ is halo mass in units of $10^{15} \hinv\msol$ and $n_{15}$
is the space density at that mass.  We show here that convolving the
mass function with a log-normal kernel of the form,
equation~(\ref{eq:Lpdf}), has an effect that is degenerate in
normalization and scatter.    

For compactness, let $x \se \lnM$, $y \se \lnL$, $a \se {\ln}L_{15}$,
then $\bar{y}(x) \se a + p \, x$ and the mass associated with a
luminosity $y$ is $\hat{x}(y) \se (y-a)/p$.  The luminosity function
is the convolution  
\begin{equation}
n(y) \, dy \ = \int dx \, n(x) \, p(y|x) \, dy ,
\end{equation}
with the kernel a Gaussian of fixed width $\sigL$
\begin{equation}
p(y|x) \, dy \ = \ \frac{1}{\sqrt{2 \pi}  \sigL} \, \exp{\left(
    -(y-\bar{y}(x))^2 / 2 \sigL^2 \right)} \, dy. 
\end{equation}

Linearly transforming the kernel, and using $\sigM  = \sigL / p$, the
luminosity function is  
\begin{equation}
n(y) \, dy \ = \ \frac{n_{15}}{\sqrt{2 \pi} \, p \, \sigM } \int dx
\exp{\left( -\gamma x - (x-\hat{x})^2/2 \sigM ^2 \right)}\,  dy 
\end{equation}

Completing the square leads to the result
\begin{equation}
n(y) \, dy \ = \ \frac{n_{15}}{p} \exp{\left(-\gamma/p [ y - (a +
    \gamma p \sigM ^2 / 2)]\right)} \, dy 
\label{eq:lfpl}
\end{equation}
which in original notation is 
\begin{equation}
n(L) \ d\lnL \ = \ \frac{n_{15}}{p} \ (L/\tilde{L}_{15})^{-\gamma/p} \ d\lnL .
\end{equation}
The result is a power-law with slope $-\gamma/p$ and with intercept 
\begin{equation}
\ln \tilde{L}_{15} \ = \ \ln L_{15} \ + \ \gamma \,p \,\sigM ^2 / 2 .
\end{equation}
Since $\tilde{L}_{15}$ is the measurable quantity (call it $C$),
then the observed luminosity function constrains the combination of
intercept and scatter 
\begin{equation}
\ln L_{15} = C - \frac{\gamma\, p \, \sigM ^2}{2}
\label{eq:lnldm}
\end{equation}
For values of $\gamma = 3.1$, the local slope of the mass function
at $10^{15}\hinv\msol$ for a concordance model, and $p = 1.6$, this
expectation is shown as the bold line in the 
lower right panel of Figure \ref{fig:mlcounts}.




\clearpage
\twocolumn




\end{document}